\documentclass[aps,prb,a4paper,twocolumn,10pt]{revtex4-2}

\usepackage{amssymb}
\usepackage{amsmath}
\usepackage{amsfonts}
\usepackage{graphicx}
\usepackage{bm}
\usepackage{textcomp}
\usepackage{physics}
\usepackage[dvipsnames]{xcolor}
\usepackage{multirow}
\usepackage{bbold}
\usepackage{comment}
\usepackage{natbib}
\usepackage{subfigure} 
\usepackage[normalem]{ulem}
\DeclareMathOperator{\sgn}{sgn}

\renewcommand{\vec}[1]{\mathbf{#1}}
\newcommand{\ve}{\vec e}
\newcommand{\vh}{\vec h}

\newcommand{\vm}{\vec m}
\newcommand{\vsigma}{\mbox{\boldmath $\sigma$}}

\newcommand{\skipthis}[1]{}

\begin{document}

\title{Long-range spin-polarized Josephson effect in ballistic S/F/S junctions with precessing magnetization}

\author{E. S. Andriyakhina}
\affiliation{Dahlem Center for Complex Quantum Systems, Fachbereich Physik, and Halle-Berlin-Regensburg Cluster of Excellence CCE, Freie Universit\"at Berlin, Arnimallee 14, 14195 Berlin}

\author{M. Mansouri}
\affiliation{Dahlem Center for Complex Quantum Systems, Fachbereich Physik, and Halle-Berlin-Regensburg Cluster of Excellence CCE, Freie Universit\"at Berlin, Arnimallee 14, 14195 Berlin}

\author{M. Breitkreiz}
\affiliation{Dahlem Center for Complex Quantum Systems, Fachbereich Physik, and Halle-Berlin-Regensburg Cluster of Excellence CCE, Freie Universit\"at Berlin, Arnimallee 14, 14195 Berlin}

\author{P. W. Brouwer}
\affiliation{Dahlem Center for Complex Quantum Systems, Fachbereich Physik, and Halle-Berlin-Regensburg Cluster of Excellence CCE, Freie Universit\"at Berlin, Arnimallee 14, 14195 Berlin}

\date{\today}

\begin{abstract}
We present a theory of ballistic N/F/S and S/F/S junctions with a uniformly precessing magnetization, which generates long-range equal-spin superconducting correlations [Takahashi {\em et al.}, Phys.\ Rev.\ Lett.\ {\bf 99}, 057003 (2007), Houzet, Phys.\ Rev.\ Lett.\ {\bf 101}, 057009 (2008)]. The non-equilibrium distribution of Andreev bound states leads to a strongly non-sinusoidal current-phase relationship for large precession angles. We derive detailed results for ballistic junctions involving partially and fully polarized ferromagnets. In the fully polarized half-metal limit, the magnetization precession switches the junction from an ``off'' state with vanishing subgap current to an ``on'' state with finite Andreev conductance and finite Josephson current.
\end{abstract}

\maketitle

\section{Introduction}

Proximity effects at the boundary between a superconductor (S) and a non-superconducting material have been extensively studied since the mid-20th century~\cite{Holm1932, Meissner1960, Josephson1962, deGennes1963, Werthamer1963}. When a superconductor is brought into contact with a normal metal (N), a region of induced superconducting coherence in the normal metal near the interface is created — a phenomenon known as the proximity effect \cite{deGennes1964,Pannetier2000}. Manifestations of the superconductor proximity effect are a suppressed density of states in the normal metal \cite{Deutscher1969, Gueron1996} and the flow of a dissipationless supercurrent through the normal metal in an S/N/S Josephson junction \cite{Kulik1970, Jackel1974}.

By tailoring the normal metal or the N/S interface, the proximity effect can be used to induce non-conventional (i.e., non-$s$-wave/spin-singlet) forms of superconductivity, thus significantly increasing the materials base beyond that of intrinsically superconducting materials. Via this principle, proximity superconductivity has been proposed as an avenue to engineer one- and two-dimensional topological superconductors \cite{Sau2009, Alicea2010, Lutchyn2010, Oreg2010, Potter2010, Duckheim2011, Qi2011, Xu2011} and as an avenue towards superconductors with odd-frequency spin-polarized Cooper pairs \cite{Bergeret2001, Kadigrobov2001, Eschrig2003, Krawiec2004, Bergeret2005, Keizer2006, Sosnin2006, Houzet2007,Braude2007, Khaire2010, Anwar2010, Sprungmann2010, Wang2010, Robinson2010, Hubler2012, Gingrich2012, Anwar2012,Banerjee2014}. Long-ranged dissipationless spin-polarized currents are key to applications in superconducting spintronics~\cite{Eschrig2015, Linder2015, Yang2021}.

If a superconductor is brought in contact with a ferromagnetic metal (F), coherent pairs of electrons with opposite spin appear in F \cite{Buzdin1982}, with spin quantization axis along the magnetization direction. They oscillate between the spin singlet and unpolarized spin triplet states \cite{Buzdin2005}, with period characterized by the difference $2\kappa = k_{\uparrow} - k_{\downarrow}$ of the Fermi momenta for the two spin projections in F. In two and three dimensions, the destructive interference of many such oscillating trajectories leads to a rapid decay of coherence away from the S/F interface. Such a decay does not occur for spin-polarized Cooper pairs, since these consist of two electrons with the same Fermi momentum in F.
Spin-polarized Cooper pairs can emerge when a conventional superconductor is placed in contact with a magnetic material with a non-collinear spin texture \cite{Bergeret2001, Houzet2007, Volkov2010}. The inhomogeneous magnetization converts spin-singlet Cooper pairs into spin-polarized pairs aligned with the (local) magnetization, thus inducing spin-polarized proximity superconductivity~\cite{Buzdin1982, Fogelstrom2000, Bergeret2001, Kadigrobov2001, Eschrig2003, Krawiec2004, Bergeret2005, Houzet2007, Volkov2010, Kupferschmidt2011, Chaou2025}.
Experimental realizations of this mechanism have been reported for magnetic multilayers with engineered non-collinearity~\cite{Khaire2010, Gingrich2012, Anwar2012, Banerjee2014}, disordered magnetic interfaces~\cite{Sprungmann2010, Wang2010}, helical magnetization profiles~\cite{Sosnin2006, Robinson2010}, and non-collinear antiferromagnets~\cite{Jeon2021, Jeon2023}. Long-range supercurrents have also been observed in a half-metallic ferromagnet (HM) \cite{Keizer2006, Anwar2010}. Since a half metal has carriers of one spin direction only, it only admits spin-polarized forms of superconductivity \cite{Eschrig2003}.

\begin{figure}[t!]
\centerline{
    \includegraphics[width=.45\textwidth]{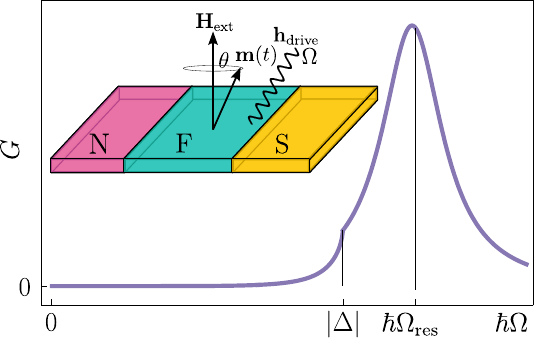}
    }
\caption{Zero-bias conductance $G(\Omega)$ of a ballistic N/HM/S junction at $T \to 0$ as a function of the frequency $\Omega$ of the ac magnetic field driving the magnetization precession. The zero-frequency conductance vanishes, because Andreev reflection is forbidden in the absence of precession. The conductance exhibits a peak for $\Omega$ equal to the ferromagnetic resonance frequency $\Omega_{\rm res}$ (which we chose to be $1.25|\Delta|$ for this plot) and has a discontinuous derivative w.r.t.\ $\Omega$ at $\hbar\Omega = |\Delta|$, where the continuum spectrum of the superconductor begins.}
\label{fig:G0_res}
\end{figure}

In pioneering works by Takahashi {\em et al}.~\cite{Takahashi2007} and Houzet~\cite{Houzet2008}, it was proposed that a precessing magnetization can also serve as a source of spin-polarized superconducting correlations. The mechanism for the generation of equal-spin Cooper pairs is similar to that in systems with spatially non-collinear magnetization, albeit relying instead on non-collinearity in time \cite{Takahashi2007, Houzet2008, Hikino2008, Braude2008, Hikino2011, Holmqvist2011, Holmqvist2012, Holmqvist2014, Kulikov2024}. In contrast to the static effect with a spatially non-uniform magnetization, where the spin-polarized component is always present, temporal non-collinearity enables dynamic control and switching of the spin-polarized superconducting component. Although supercurrents facilitated by magnetization precession have not been realized experimentally to the best of our knowledge, several experiments have reported signatures of a coupling between magnetization dynamics and superconductivity \cite{Jeon2018,Li2018,Chan2023,Li2025}. 

In this work, we consider spin-polarized proximity superconductivity in a ballistic junction involving a ferromagnetic metal with a uniformly precessing magnetization. This limit is particularly relevant for vertical Josephson junctions (as opposed to lateral junctions), where transport occurs along the growth direction of the heterostructure. In such samples, the mean free path can be comparable to the sample thickness, which is typically on the order of several nanometers. As a result, transport may be close to ballistic. We relate the Josephson current $I$ to the dependence of Andreev bound state energies on the phase difference $\phi$ across the junction. This approach is especially useful in the limit of short junctions (junction length $L$ much shorter than the superconductor coherence length $\xi$), because such junctions contain only few Andreev levels, which dominate the supercurrent response. In addition to junctions with two superconducting contacts, we also consider junctions with one superconducting and one normal contact. In such junctions the dependence of  the conductance $G$ on voltage, precession frequency, and temperature gives key information on the equal-spin Andreev reflection facilitating the spin-polarized proximity effect in these junctions. Our work complements previous theories of the Josephson effect with a precessing magnetization, which focused on the diffusive limit and tunneling barriers separating superconducting and normal regions \cite{Takahashi2007,Houzet2008}.

In a conventional Josephson junction, Andreev bound states occupy the full subgap range $0 \le E \le |\Delta|$, while the spectrum is continuous only for excitation energies $E > |\Delta|$. In the ``short-junction limit'', the Josephson current $I$ is then determined entirely by the bound states \cite{beenakker1991d, Beenakker1992}. For a junction with precessing magnetization, this picture changes qualitatively. Bound states exist only in the reduced energy range $0 \le E < |\Delta| - \hbar |\Omega|/2$, so that part of the subgap region is replaced by a continuous spectrum. As a result, unlike in the conventional case, the continuous part of the spectrum also contributes to the Josephson current. At the same time, the discrete Andreev bound states still leave a strong imprint on the current, which changes abruptly whenever a level crosses $\hbar \Omega/2$ or merges with the continuum.
This interplay between discrete and continuous contributions gives rise to a strongly non-sinusoidal current-phase relation for large tilt angles $\theta$ \cite{Holmqvist2012} and to abrupt jumps in the Josephson current as the junction length $L$ is increased.

For small amplitudes of the precessing magnetization, which is measured by the tilt angle $\theta$, see Fig.\ \ref{fig:G0_res}, the Josephson current $I$ (in an S/F/S junction) and the conductance $G$ (in an N/F/S junction) are proportional to $\theta^2$. As pointed out by Takahashi {\em et al.}, this dependence on the amplitude of the precessing magnetization implies a resonant enhancement of the critical current and the conductance if the precession frequency $\Omega$ matches the ferromagnetic resonance frequency $\Omega_{\rm res}$ \cite{Takahashi2007}. This effect, which is a unique signature of microwave-controlled switching of the spin-polarized Josephson effect, is illustrated in Fig.\ \ref{fig:G0_res}.

The remainder of this article is organized as follows: We introduce the scattering formalism for a ballistic, one-dimensional junction in Sec.\ \ref{sec:2}. In Sec.\ \ref{Sec:SHMS} we specialize to the case that the magnetic element F is fully spin-polarized. Precession-induced effects are particularly striking in such a half-metallic junction, because there is no proximity superconductivity without precession. Junctions containing a partially polarized ferromagnetic  element F are considered in Sec.\ \ref{Sec:SFMS}. In this case, spin-polarized and opposite-spin proximity superconductivity coexist. A key difference between the two forms of superconductivity, is that the latter oscillates starkly with junction length $L$ with zero average, whereas the former has a component that remains finite after averaging over $L$. 
As a result, in the partially polarized case, the opposite-spin proximity effect is short-ranged in two and three dimensions, whereas the equal-spin proximity effect is long-ranged, as we discuss in Sec.\ \ref{Sec:2D:3D}.
We conclude with a brief discussion of our results in Sec.\ \ref{Sec:Discussion}.

\section{One-dimensional junctions: formalism}
\label{sec:2}

We first consider a one-dimensional junction, consisting of a central ferromagnetic or half-metallic layer F of length $L$ and one superconducting and one normal-metal contact N or two superconducting contacts S with a phase difference $\phi$. We choose coordinates such that the magnetic part of the junction is for $-L/2 < x < L/2$ and the two leads are at $x < -L/2$ and $x > L/2$, as shown schematically in Fig.\ \ref{fig:setup}a and b. 

The junction is described by the second-quantized Hamiltonian
\begin{align}
  \hat H(t) =&\, \int dx\, \sum_{\sigma,\sigma'}
  \hat{\psi}_{\sigma}^\dag (x) {\cal H}_{\sigma,\sigma'}(x,t) \hat{\psi}_{\sigma'} (x) \nonumber \\
   &\, \mbox{}
  + \int dx [\hat{\psi}_{\uparrow}^\dag (x) \hat{\psi}_{\downarrow}^\dag (x) \Delta(x) + \mbox{h.c.}],  \label{eq:H}
\end{align}
where the superconducting order parameter $\Delta(x)$ is non-zero in the superconducting regions only,
\begin{equation}
    \Delta(x) = |\Delta| e^{i\phi} \Theta(x-L/2)
\end{equation}
for the case of one superconducting contact (Fig.\ \ref{fig:setup}a) and
\begin{equation} \label{eq:Delta(x)}
    \Delta (x) =|\Delta| e^{i \sgn(x) \phi/2} \Theta(|x| - L/2)
\end{equation}
for the case of two superconducting contacts (Fig.\ \ref{fig:setup}b).
In the leads, the normal-state Hamiltonian ${\cal H}$ is time-independent and takes the simple form
\begin{equation}
  {\cal H}(x) = \left( -\frac{\hbar^2}{2 m} \frac{\partial^2}{\partial x^2} - \mu \right) \sigma_0, \label{eq:Hnormal}
\end{equation}
where $\sigma_0$ is the identity matrix for the spin degrees of freedom. In the magnetic part of the junction, $-L/2 < x < L/2$, the Hamiltonian ${\cal H}$ is time-dependent because of the precessing magnetization. It reads
\begin{equation}
  {\cal H}(x,t) = \left( -\frac{\hbar^2}{2 m} \frac{\partial^2}{\partial x^2}
  + V_{\rm F} 
  - \mu \right) \sigma_0 - h_{\rm F} \vm(t) \cdot \vsigma, \label{eq:HF}
\end{equation}
where $\vsigma=(\sigma_x,\sigma_y,\sigma_z)$, $V_{\rm F}$ is a potential offset, $h_{\rm F}$ the exchange field, and $\vm(t)$ is the time-dependent magnetization direction, which precesses at angular frequency $\Omega$ around the $z$ axis,
\begin{equation}
  \vm(t) = (\sin \theta \cos \Omega t,\ \sin \theta \sin \Omega t,\ \cos \theta). \label{eq:mt}
\end{equation}

\subsection{Rotating frame}

\begin{figure}[t!]
\centerline{
    \includegraphics[width=0.45\textwidth]{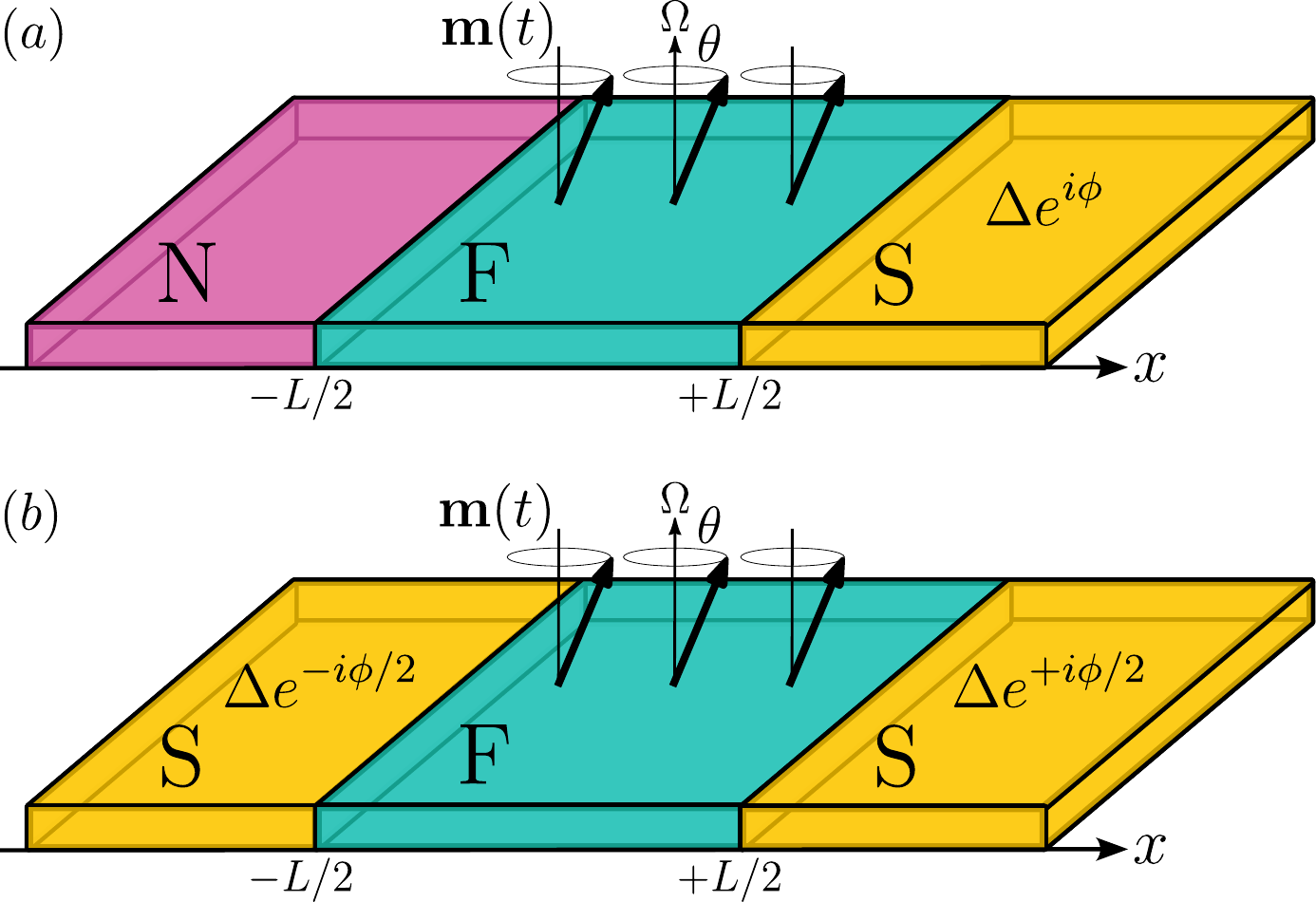}
    }
\caption{Geometry of the $(a)$ N/F/S and $(b)$ S/F/S junctions. In both panels, the region $-L/2 < x < L/2$ contains a magnet with precessing magnetization $\vm(t)$ with precession frequency $\Omega$ and tilt angle $\theta$. For the S/F/S junction (b), the two superconducting contacts have phase difference $\phi$. 
}
\label{fig:setup}
\end{figure}

The time-dependence of the Hamiltonian (\ref{eq:H}) 
can be removed by switching to a rotating reference frame for the spin degree of freedom. Such a shift of reference frame is accomplished by the transformation~\cite{Takahashi2007, Houzet2008,Teber2010,Holmqvist2011,Holmqvist2012,Holmqvist2014}
\begin{align}
  \label{eq:U_t}
  \hat{\tilde{\psi}}_\sigma(x) =&\, e^{\frac{1}{2} i \sigma \Omega t} \hat{\psi}_{\sigma}(x),
\end{align}
where we use a tilde $\tilde{\cdot}$ to indicate operators in the rotating frame. In the rotating frame, the Hamiltonian is time independent,
\begin{align}
  \hat{\tilde{H}} =&\, \int dx \sum_{\sigma,\sigma'} \hat{\tilde{\psi}}_{\sigma}^\dag (x) \tilde {\cal H}_{\sigma,\sigma'}(x) \hat{\tilde{\psi}}_{\sigma'} (x)
  \nonumber \\ &\, \mbox{}
    + \int dx  [\hat{\tilde{\psi}}_{\downarrow}^\dag (x) \hat{\tilde{\psi}}_{\uparrow}^\dag (x) \Delta(x) + \mbox{h.c.}],
  \label{eq:tildeH}
\end{align}
where
\begin{equation}
  \tilde {\cal H}(x) = {\cal H}(x,0) - \frac{1}{2} \hbar \Omega \sigma_z.
  \label{eq:tildeHnormal}
\end{equation}
The spin-singlet superconducting order parameter $\Delta(x)$ is not affected by the transformation.

The transformation to the rotating frame exchanges the time dependence of the Hamiltonian for an additional exchange field $\tfrac{1}{2} \hbar \Omega$ along the $z$ direction. Hence, if the tilt angle $\theta \neq 0$, the Hamiltonian (\ref{eq:tildeH}) has non-collinear exchange fields: In the leads, the effective exchange field is $\tfrac{1}{2}\hbar \Omega \ve_z$, whereas in F the effective exchange field
\begin{equation}
  \vh_{\rm eff} = h_{\rm F} \vm(0) + \frac{1}{2} \hbar \Omega \ve_z
  \label{eq:heff}
\end{equation}
also has a component perpendicular to $\ve_z$. As discussed in the introduction, the presence of non-collinear exchange fields is known to be a route to transform spin singlet superconducting correlations into spin-polarized correlations \cite{Bergeret2001, Houzet2007, Volkov2010}. 

We write the second-quantized Hamiltonian (\ref{eq:tildeH}) in terms of the $4 \times 4$ Bogoliubov-de Gennes Hamiltonian ${\cal H}_{\rm BdG}$,
\begin{equation}
  \hat{\tilde H} = \frac{1}{2} \int dx \hat{\tilde{\Psi}}(x)^{\dagger} \tilde {\cal H}_{\rm BdG}(x) \hat{\tilde{\Psi}}(x),
\end{equation}
where $\hat{\tilde{\Psi}}(x) = (\hat{\tilde{\psi}}_{\uparrow}(x),\hat{\tilde{\psi}}_{\downarrow}(x),\hat{\tilde{\psi}}_{\uparrow}^{\dagger}(x),\hat{\tilde{\psi}}_{\downarrow}^{\dagger}(x))^{\rm T}$,
\begin{equation}
  \tilde {\cal H}_{\rm BdG}(x) = \begin{pmatrix} \tilde {\cal H}(x) & -i \sigma_y \Delta(x) \\ i \sigma_y \Delta(x)^* & - \tilde {\cal H}(x)^* \end{pmatrix},
  \label{eq:tildeHBdG}
\end{equation}
and we neglected an additive constant that does not enter into the calculation of the current.

To diagonalize the rotating-frame Hamiltonian, we introduce the eigenvectors of the Bogoliubov-de Gennes Hamiltonian (\ref{eq:tildeHBdG}),
\begin{equation}
  \tilde{{\cal H}}_{\rm BdG} \Psi_{\alpha}(x)
  =
  \tilde E_{\alpha} \Psi_{\alpha}(x),\ \ 
  \Psi_{\alpha}(x) = \begin{pmatrix} u_{\alpha;\uparrow}(x) \\
  u_{\alpha;\downarrow}(x) \\ v_{\alpha;\uparrow}(x) \\ v_{\alpha;\downarrow}(x) \end{pmatrix},
  \label{eq:HBdGEq}
\end{equation}
at positive eigenvalue $\tilde E_{\alpha} > 0$. 

In the case of two superconducting contacts (geometry of Fig.\ \ref{fig:setup}b), the spectrum of ${\cal H}_{\rm BdG}$ is discrete for $0 \le \tilde{E}_{\alpha} < |\Delta| - \hbar |\Omega|/2$. In the discrete part of the spectrum, the eigenstates $\Psi_{\alpha}(x)$ of ${\cal H}_{\rm BdG}$ are Andreev bound states~\cite{Andreev1964}, which have a normalized four-component wavefunction,
\begin{equation}
  \sum_{\sigma'} \int dx \left( |u_{\alpha;\sigma'}(x)|^2 + |v_{\alpha;\sigma'}(x)|^2 \right) = 1.
  \label{eq:normalization}
\end{equation}

The spectrum of ${\cal H}_{\rm BdG}$ is continuous for eigenvalues $\tilde{E}_{\alpha} > |\Delta| - \hbar |\Omega|/2$. If there is one normal-metal contact (geometry of Fig.\ \ref{fig:setup}a), the spectrum of ${\cal H}_{\rm BdG}$ is continuous for all eigenvalues. 
In the continuous part of the spectrum, we choose the four-component wavefunction $\Psi_{\alpha}(x)$ in the form of a scattering state with normalized quasiparticle flux. In this part of the spectrum the eigenstate index $\alpha = (\tilde{E},\beta,\tau,\sigma)$ is a composite index, where $\tilde E$ is the eigenvalue of ${\tilde {\cal H}}_{\rm BdG}$ and the three labels $\beta$, $\tau$, and $\sigma$ indicate, whether the incoming-wave part of the scattering state $\Psi_{\alpha}(x)$ is incident from the right ($\beta =1$) or from the left ($\beta = -1$), whether it is predominantly electron-like ($\tau = 1$) or hole-like ($\tau = -1$), and whether it corresponds to an excitation with spin $\sigma = 1$ or $\sigma = -1$ along the $z$ axis. Explicitly, the {\em incoming-wave part} of the wavefunction $\Psi_{\alpha}(x)$ with $\alpha = (\tilde E,\beta,\tau,\sigma)$ is \cite{beenakker1991d},
\begin{align}
  \label{eq:Psi}
  \Psi_{\alpha}(x) =&\,
  \frac{e^{- i \tau q_{\tau \sigma} (\beta x - \tfrac{1}{2}L)}}{\sqrt{v_{\tau \sigma}}} \Phi_{\alpha},\ \
  \beta x > \tfrac{1}{2} L,
\end{align}
where $q_{\tau \sigma}$ is the positive solution of
\begin{align}
  \frac{(\hbar q_{\tau \sigma})^2}{2 m}  =&\, \mu + \tau \sqrt{(\tilde E + \tfrac{1}{2} \hbar \sigma \Omega)^2 - |\Delta|^2},
\end{align}
$v_{\tau \sigma} = |\hbar d q_{\tau \sigma}/d\tilde E|^{-1}$ is the velocity, and the four-component spinors $\Phi_{\alpha}$ are
\begin{align}
    \Phi_{\tilde E,\beta\tau\uparrow} =&\,
    \frac{1}{\sqrt{2 \cos \eta_{+}}}
  \begin{pmatrix} e^{\tfrac{i}{2} (\beta \phi + \tau \eta_+)} \\ 0 \\ 
  0 \\ e^{-\tfrac{i}{2}  (\beta \phi + \tau \eta_+)} \end{pmatrix}, \nonumber \\
  \Phi_{\tilde E,\beta\tau\downarrow} =&\,
  \frac{1}{\sqrt{2 \cos \eta_{-}}}
  \begin{pmatrix} 0 \\ e^{\tfrac{i}{2} (\beta \phi + \tau \eta_-)} 
  \\ - e^{-\tfrac{i}{2} (\beta \phi + \tau \eta_-)} \\ 0\end{pmatrix},
\end{align}
with
\begin{equation}
  \eta_{\sigma} = -i\, \mbox{arcosh}\frac{\tilde E + \tfrac{1}{2} \hbar \sigma \Omega}{|\Delta|}.
  \label{eq:etacontinuous}
\end{equation}
The limit $|\Delta| \to 0$ must be taken for scattering states incident from the normal contact in case of the geometry of Fig.\ \ref{fig:setup}a. 
In the range $\tilde E < |\Delta| - \hbar |\Omega|/2$ only scattering states incident from the normal contact exist for the geometry of Fig.\ \ref{fig:setup}a, so that $\beta = -1$.
In the range $|\Delta| - \hbar |\Omega|/2 < \tilde{E} < |\Delta| + \hbar |\Omega|/2$ only scattering states with $\sigma = \sgn \Omega$ exist in the superconducting contact.

Two remarks are in order: (i) If $\theta \neq 0$ it is not possible to assign a unique spin direction to the scattering state $\Psi_{\alpha}$ as a whole, because of the non-collinear exchange fields in the rotating-frame Hamiltonian $\tilde {\cal H}$ of Eq.\ (\ref{eq:tildeHnormal}). However, the incoming part of the Bogoliubov scattering states does have a well-defined spin along $z$, see Eq.\ (\ref{eq:Psi}), which is what we use to label the scattering state. (ii) If $\theta = 0$ the sectors of the Hilbert space with spin up and spin down decouple. In this case in a junction with two superconducting contacts, Andreev bound states with discrete energies $E_{\alpha}$ and spin $-\sgn(\Omega)$ coexist with the continuous-spectrum states with spin $\sigma = \sgn(\Omega)$ in the energy range $|\Delta| - \hbar |\Omega|/2 < \tilde{E} < |\Delta| + \hbar |\Omega|/2$. 

In the non-rotating reference frame, scattering states incoming from the leads at energy $E$ have the equilibrium occupation $f(E)$, where $f$ is the Fermi-Dirac function. Since the transition to the rotating frame involves a shift of the normal-state Hamiltonian by $-\tfrac{1}{2} \sigma \hbar \Omega$
for $\sigma = \uparrow$ ($+$) and $\sigma = \downarrow$ ($-$), respectively, the rotating-frame energy $\tilde E$ of these states is 
$\tilde E = E - \tfrac{1}{2} \sigma \hbar \Omega$,
see Eq.\ (\ref{eq:tildeHnormal}).
Hence, the scattering state $\alpha = (\tilde E,\beta,\tau,\sigma)$ in the rotating frame has the occupation
\begin{equation}
  \tilde n_{\alpha} = f_{\sigma}(\tilde E), \label{eq:n_scattering}
\end{equation}
where we abbreviate
\begin{equation}
  f_{\sigma}(\tilde E) = f(\tilde E + \tfrac{1}{2} \hbar \sigma \Omega).
\end{equation}
To determine the occupation of the Andreev bound states in the discrete part of the spectrum of an S/F/S junction, we imagine that F is (infinitesimally) tunnel-coupled to a normal reservoir, which is in thermal equilibrium and extends homogeneously along $x$. In the normal reservoir, quasiparticle excitations at spin $\sigma$ have the occupation $f_{\sigma}(\tilde E)$ in the rotating frame. We choose the tunneling rate for spin-$\sigma$ quasiparticles to and from the fictitious reservoir to be proportional to
\begin{equation}
    \rho_{\alpha\sigma} = \int_{-L/2}^{L/2} dx (|u_{\alpha;\sigma}(x)|^2 + |v_{\alpha;-\sigma}(x)|^2),
    \label{eq:rhoalpha}
\end{equation}
and that the steady-state occupation of the Andreev bound state $\alpha$ is
\begin{equation}
  \label{eq:nalpha}
  \tilde n_{\alpha} = 
  \frac{\rho_{\alpha;\uparrow} f_{\uparrow}(\tilde E_{\alpha}) +
  \rho_{\alpha;\downarrow} f_{\downarrow}(\tilde E_{\alpha})}{\rho_{\alpha;\uparrow} + \rho_{\alpha;\downarrow}}.
\end{equation}
Hence, the population of the Andreev bound states/Bogoliubov scattering state $\alpha$ differs from the thermal equilibrium distribution in both the discrete and the continuous parts of the spectrum. 

We remark that the choice (\ref{eq:rhoalpha}) for weight factors $\rho_{\alpha\sigma}$ is not unique, because it depends on the type of relaxation processes that determine the steady-state population $n_{\alpha}$ in the rotating frame. These relaxation processes are not included in the simple mean-field Hamiltonian (\ref{eq:H}). Microscopically, the steady-state population is determined by inelastic scattering processes that couple the occupation of the Andreev bound states to that of highly excited quasiparticles in the superconducting leads. These transitions are mediated by interactions with a bosonic bath, {\em e.g.}, phonons. The tunnel coupling to a normal reservoir used to construct Eq.\ (\ref{eq:n_scattering}) serves as a proxy for such inelastic processes in a single-particle framework, similar to B\"uttiker's voltage-probe model for dephasing \cite{buettiker1986}. The choice of the integration limits in Eq.\ (\ref{eq:nalpha}) assumes that relaxation takes place in the central magnetic element of the junction only. For comparison, a model in which relaxation takes place equally inside the superconductor and the magnet would use the full support of the Andreev bound states to obtain the weight factors $\rho_{\alpha\sigma}$. We refer to Refs.\ \cite{Holmqvist2011,Teber2010,Holmqvist2012,Holmqvist2014} for a discussion of the population of the Andreev bound states in the rotating frame in the limit of a tunnel junction connecting the two superconducting contacts.

\subsection{Scattering approach}
\label{sec:2b}

\begin{figure}[t!]
\centerline{
    \includegraphics[width=.45\textwidth]{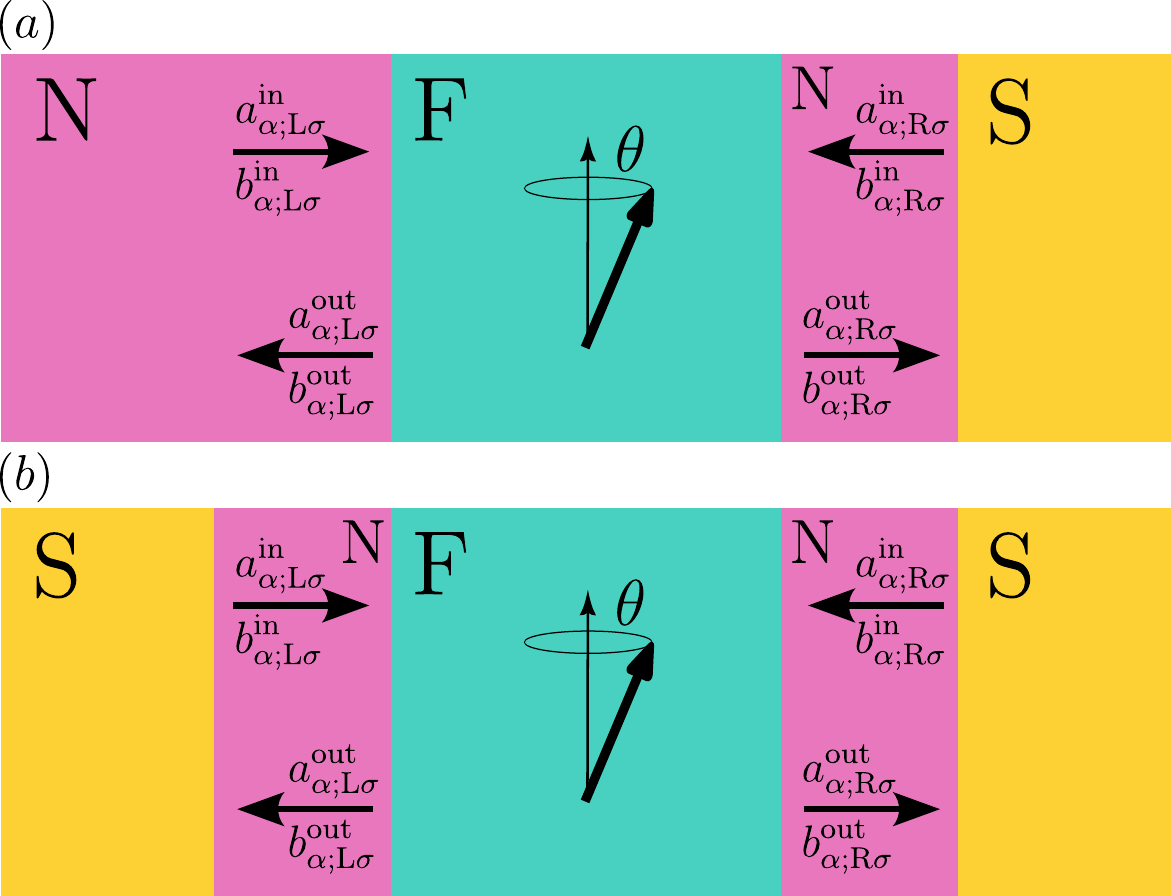}
    }
\caption{System geometry with additional ideal leads between the ferromagnet and the left and right contacts for the case of one normal contact and one superconducting contact $(a)$ and two superconducting contacts $(b)$. The amplitudes $a^{\rm in}_{\alpha,\beta'\sigma}$ and $a^{\rm out}_{\alpha,\beta'\sigma}$ ($b^{\rm in}_{\alpha,\beta'\sigma}$ and $b^{\rm out}_{\alpha,\beta'\sigma}$), with $\beta' = {\rm L}, {\rm R}$, refer to electron-like (hole-like) states with spin $\sigma$ propagating toward and away from the ferromagnet, respectively.}
\label{fig:modes}
\end{figure}

To find the Andreev bound states (in the discrete part of the spectrum) and the Bogoliubov scattering states (in the continuous part of the spectrum), we follow a scattering approach. We insert ideal leads between F and the leads, see Fig. \ref{fig:modes}. The length of the ideal leads is sent to zero at the end of the calculation.

In the ideal lead at the interface with the right or left contact ($\beta' = 1$, $-1$ or ${\rm R}$, ${\rm L}$, respectively), the four-component wavefunction $\Psi_{\alpha}(x)$ in the BdG formalism can be written as a sum of components moving towards F (``in'') and away from F (``out''),
\begin{align}
  \Psi_{\alpha}(x) =&\, \frac{e^{-i k_0 (\beta' x - \tfrac{L}{2})}}{\sqrt{v}}
  \begin{pmatrix}
      a^{\rm in}_{\alpha,\beta' \uparrow} \\
      a^{\rm in}_{\alpha,\beta' \downarrow} \\
      b^{\rm out}_{\alpha,\beta' \uparrow} \\
      b^{\rm out}_{\alpha,\beta' \downarrow}
  \end{pmatrix}
  \nonumber \\ &\, \mbox{}
  + \frac{e^{i k_0 (\beta' x - \tfrac{L}{2})}}{\sqrt{v}} 
  \begin{pmatrix}
      a^{\rm out}_{\alpha,\beta' \uparrow} \\
      a^{\rm out}_{\alpha,\beta' \downarrow} \\
      b^{\rm in}_{\alpha,\beta' \uparrow} \\
      b^{\rm in}_{\alpha,\beta' \downarrow}
  \end{pmatrix},
  \label{eq:uvR}
\end{align}
where $\hbar^2 k_0^2/2m = \mu$ and $v = \hbar k_0/m$. (For the wavefunctions in the ideal leads, we neglect $|\tilde E|$ and $\hbar |\Omega|$ with respect to the chemical potential $\mu$.) The amplitudes $a_{\alpha,\beta'\sigma'}^{{\rm in/out}}$ and $b_{\alpha,\beta'\sigma'}^{{\rm in/out}}$ are flux normalized, {\em i.e.}, they have dimension $(\mbox{time})^{-1/2}$.

The total current through the junction is
\begin{equation}
  I = I_{\rm R} = -I_{\rm L},
    \label{eq:ItotalRL}
\end{equation}
with
\begin{align}
  \label{eq:Itotal}
  I_{\beta'} =&\, -\beta' e \sum_{\alpha} \sum_{\sigma'}
   \left[ (|b^{{\rm in}}_{\alpha;\beta'\sigma'}|^2
  - |b^{{\rm out}}_{\alpha;\beta'\sigma'}|^2) (1 - \tilde n_{\alpha})
  \right. \nonumber \\ &\, \left. \mbox{}
  + (|a^{{\rm in}}_{\alpha;\beta'\sigma'}|^2
  - |a^{{\rm out}}_{\alpha;\beta'\sigma'}|^2) \tilde n_{\alpha} \right],
\end{align}
where we recall that in the continuous part of the spectrum $\alpha = (\tilde E,\beta,\tau,\sigma)$ is a composite index, so that the symbol $\Sigma_{\alpha} \ldots$ should be understood as $\sum_{\beta, \sigma, \tau} \int d\tilde{E} \ldots$. For the discrete part of the spectrum, one may make use of the identity \cite{beenakker1991d}
\begin{align}
  \label{eq:IEdiscrete}  
  \frac{\partial \tilde E_{\alpha}}{\partial \phi} =&\,
  - \beta' \hbar  \sum_{\sigma'}
  \left[ (|b^{{\rm in}}_{\alpha;\beta'\sigma'}|^2
    - |b^{{\rm out}}_{\alpha;\beta'\sigma'}|^2)
    \right] \nonumber \\ =&\,
  \beta' \hbar \sum_{\sigma'}
  \left[ (|a^{{\rm in}}_{\alpha;\beta'\sigma'}|^2
    - |a^{{\rm out}}_{\alpha;\beta'\sigma'}|^2) 
    \right]
\end{align}
to simplify the calculation of the current. 

In the scattering approach, the solution of the Bogoliubov-de Gennes equation (\ref{eq:HBdGEq}) is written in terms of matrix equations relating the coefficients in Eq.\ (\ref{eq:uvR}). For the central F part, the electron ($u$) components are related as
\begin{align}
  \label{eq:SF}
  \begin{pmatrix}
    a_{\alpha;{\rm L}\uparrow}^{{\rm out}} \\
    a_{\alpha;{\rm L}\downarrow}^{{\rm out}}\\
    a_{\alpha;{\rm R}\uparrow}^{{\rm out}} \\
    a_{\alpha;{\rm R}\downarrow}^{{\rm out}}
  \end{pmatrix}
  =&\, S_{\rm F}(\tilde E)
  \begin{pmatrix}
    a_{\alpha;{\rm L}\uparrow}^{{\rm in}} \\
    a_{\alpha;{\rm L}\downarrow}^{{\rm in}}\\
    a_{\alpha;{\rm R}\uparrow}^{{\rm in}} \\
    a_{\alpha;{\rm R}\downarrow}^{{\rm in}}
  \end{pmatrix},
\end{align}
where the scattering matrix $S_{\rm F}(\tilde E)$ of the central F region is of the form
\begin{equation}
  S_{\rm F}(\tilde E) = \sum_{\pm} \begin{pmatrix} r_{\pm}(\tilde E) &
    t'_{\pm}(\tilde E) \\ t_{\pm}(\tilde E) & r_{\pm}'(\tilde E) \end{pmatrix}
  \otimes P_{\pm},
  \label{eq:Sproj}
\end{equation}
with 
\begin{equation}
  P_{\pm} = \frac{1}{2}(\sigma_0 \pm \tilde \vm_{\rm eff} \cdot \vsigma)
  \label{eq:Ppm}
\end{equation}
the projector on spin parallel ($+$) or antiparallel ($-$) to $\tilde \vm_{\rm eff} = \vh_{\rm eff}/|\vh_{\rm eff}|$, which is the direction of the effective exchange field in F, see Eq.\ (\ref{eq:heff}).
For the Hamiltonian (\ref{eq:HF}), the reflection and transmission amplitudes read
\begin{align}
  r_{\pm}(\tilde E) =&\, r'_{\pm}(\tilde E) \nonumber \\ =&\,
  \frac{(k^2 - k_{\pm}^2) \sin k_{\pm} L}{(k^2 + k_{\pm}^2) \sin k_{\pm} L + 2 i k k_{\pm} \cos k_{\pm} L},  \label{eq:r} \\
  t_{\pm}(\tilde E) =&\, t'_{\pm}(\tilde E) \nonumber \\ =&\,
  \frac{2 i k k_{\pm}}{(k^2 + k_{\pm}^2) \sin k_{\pm} L + 2 i k k_{\pm} \cos k_{\pm} L}, \label{eq:t}
\end{align}
where $\hbar^2 k^2/2m = \mu + \tilde E$ and $\hbar^2 k_{\pm}^2/2 m = \mu + \tilde E - V_{\rm F} \pm |h_{\rm eff}|$. For the hole ($v$) components, the wavefunction amplitudes are related by the complex conjugate of the scattering matrix,
\begin{equation}
  \label{eq:SFh}
  \begin{pmatrix}
    b_{\alpha;{\rm L}\uparrow}^{{\rm out}} \\
    b_{\alpha;{\rm L}\downarrow}^{{\rm out}}\\
    b_{\alpha;{\rm R}\uparrow}^{{\rm out}} \\
    b_{\alpha;{\rm R}\downarrow}^{{\rm out}}
  \end{pmatrix}
  = S_{\rm F}(-\tilde E)^*
  \begin{pmatrix}
    b_{\alpha;{\rm L}\uparrow}^{{\rm in}} \\
    b_{\alpha;{\rm L}\downarrow}^{{\rm in}}\\
    b_{\alpha;{\rm R}\uparrow}^{{\rm in}} \\
    b_{\alpha;{\rm R}\downarrow}^{{\rm in}}
  \end{pmatrix}.
\end{equation}

For the scattering relations at the interfaces with the superconductors, we discuss the discrete and continuous parts of the spectrum separately. For the discrete part of the spectrum, $0 \le \tilde E < |\Delta| - (1/2) \hbar |\Omega|$, one has
\begin{align}
  \label{eq:SS}
  a_{\alpha;\beta'\sigma'}^{{\rm in}} =&\, \sigma' e^{-i \eta_{\sigma'} + i \beta \phi/2}
  b_{\alpha;\beta'-\sigma'}^{{\rm out}}, \nonumber \\
  b_{\alpha;\beta'-\sigma'}^{{\rm in}} =&\, \sigma' e^{-i\eta_{\sigma'} - i \beta \phi/2}
  a_{\alpha;\beta'\sigma'}^{{\rm out}},
\end{align}
where
\begin{align}
  \eta_{\sigma} = \arccos\frac{\tilde E + \tfrac{1}{2} \hbar \sigma \Omega}{|\Delta|}
\end{align}
generalizes Eq.\ (\ref{eq:etacontinuous}) to arguments $|\tilde E + (1/2) \sigma \hbar \Omega| < |\Delta|$.
The combination of Eqs.\ (\ref{eq:SF}) and (\ref{eq:SS}) allows nontrivial solutions for specific values of the Andreev bound state energy $\tilde E_{\alpha}$ only. To find the Andreev bound state wavefunction, we also impose the normalization condition (\ref{eq:normalization}).
In the continuous part of the spectrum, Eqs.\ (\ref{eq:SS}) are replaced by 
\begin{align}
  \label{eq:SScontinuous}
  a_{\alpha;\beta'\sigma'}^{{\rm in}} =&\, \sigma' e^{-i \eta_{\sigma'} + i \beta \phi/2}
  b_{\alpha;\beta'-\sigma'}^{{\rm out}}
  + s_{\beta\tau\sigma} \delta_{\beta\beta'}\delta_{\tau{\rm e}} \delta_{\sigma\sigma'}, \nonumber \\
  b_{\alpha;\beta'-\sigma'}^{{\rm in}} =&\, \sigma' e^{-i\eta_{\sigma'} - i \beta \phi/2}
  a_{\alpha;\beta'\sigma'}^{{\rm out}} + s_{\beta\tau\sigma} \delta_{\beta\beta'}\delta_{\tau{\rm h}} \delta_{\sigma\sigma'},
\end{align}
where
\begin{equation}
  s_{\beta\tau\sigma} = e^{\tfrac{i}{2} \beta \tau \phi} \sqrt{1 - |e^{-i \eta_{\sigma}}|^2}.
\end{equation}
Equations (\ref{eq:SF}) and (\ref{eq:SScontinuous}) have a unique nontrivial solution for all $\tilde E$ in the continuous part of the spectrum. The choice of the source terms in Eqs.\ (\ref{eq:SScontinuous}) ensures that this solution is normalized to unit incoming quasiparticle flux in the left or right superconducting lead.

\section{N/HM/S and S/HM/S junctions}
\label{Sec:SHMS}

We begin by considering one-dimensional junctions with a fully polarized half-metallic magnet (HM) with precessing magnetization. In a half metal there are charge carriers of only one spin direction. This is the most illustrative scenario for the mechanism responsible for the generation of spin-polarized superconductivity, since no subgap current can flow through the junction without precessing magnetization \cite{Buzdin2005}
(except for a residual contribution associated with the emission of magnons upon Andreev reflection at the HM/S interface \cite{Tkachov2001}). 

To simplify the expressions for the scattering matrix $S_{\rm F}$ of Eq.\ (\ref{eq:Sproj}), we set $|h_{\rm eff}| = V_{\rm F} \gg \mu$ and we make the Andreev approximation $|\Delta| \ll \mu$ so that
\begin{align}
  \label{eq:rHsimpl}
  & r_{+}(\tilde E) = 0,\ \
  && t_{+}(\tilde E) = e^{i k_0 L + i \tilde E L/\hbar v_0}, \nonumber \\
  & r_{-}(\tilde E) = -1,\ \
  && t_{-}(\tilde E) = 0,
\end{align}
with $k_0 = \sqrt{2 m \mu/\hbar^2}$ and $v_0 = \hbar k_0/m$.

\subsection{N/HM/S junction \label{subsec:SHMS:1contact}}

We first consider a junction with one normal-metal and one superconducting contact, see Fig.\ \ref{fig:setup}a. To find the conductance of the junction and to characterize the induced superconductivity, we calculate the amplitudes $r_{{\rm h}s,{\rm e}s'}(\tilde E) = r_{{\rm e}s,{\rm h}s'}(-\tilde E)^*$ and $r_{{\rm e}s,{\rm e}s'}(\tilde E) = r_{{\rm h}s,{\rm h}s'}(-\tilde E)^*$ for Andreev reflection and normal reflection in the rotating frame, respectively. A simple calculation using Eqs.\ (\ref{eq:SF})--(\ref{eq:SScontinuous}) gives the equal-spin amplitudes
\begin{align}\label{eq:r_HM}
  r_{{\rm h}+,{\rm e}+}(\tilde E) =&\, 
  -2 i e^{2 i \tilde E L/\hbar v_0 - i \phi}
  \frac{\sin \eta' \cos \bar \eta \sin \theta}{e^{2 i \bar \eta} + \sin^2 \eta' \sin^2 \theta}, \nonumber \\
  r_{{\rm h}-,{\rm e}-}(\tilde E) =&\,
  0,\nonumber \\
  r_{{\rm e}+,{\rm e}+}(\tilde E) =&\, e^{2 i(k_{\rm _0} + \tfrac{\tilde E}{\hbar v_0}) L} \frac{(\sin \eta' \cos \theta + i \cos \eta')^2}{e^{2 i \bar \eta} + \sin^2 \eta' \sin^2 \theta}, \nonumber \\
  r_{{\rm e}-,{\rm e}-}(\tilde E) =&\, -1,
\end{align}
where the labels $\pm$ refer to the spin projection onto the magnetization direction $\tilde \vm_{\rm eff}$ and we abbreviated
\begin{equation}
    \bar \eta = \frac{\eta_+ + \eta_-}{2}, \quad \eta' = \frac{\eta_+ - \eta_-}{2}.
\end{equation}
The opposite-spin amplitudes all vanish, $r_{\tau -, \tau'+}(\tilde E) = r_{\tau +,\tau'-}(\tilde E) = 0$, $\tau$, $\tau' \in \{{\rm e},{\rm h}\}$. Scattering amplitudes for spin labels $\sigma$, $\sigma' = \uparrow$, $\downarrow$ defined with respect to the $z$ axis are (compare with Eq.\ (\ref{eq:Sproj}))
\begin{equation}
  \begin{pmatrix} r_{\tau \uparrow,\tau' \uparrow}(\tilde E) &
r_{\tau \uparrow,\tau' \downarrow}(\tilde E) \\ r_{\tau \downarrow,\tau' \uparrow}(\tilde E) & r_{\tau \downarrow,\tau' \downarrow}(\tilde E) \end{pmatrix}
  = \sum_{s, s' = \pm}
    P_{s} r_{\tau s,\tau' s'}(\tilde E) P_{s'},
\end{equation}
with the projector $P_{\pm}$ given by Eq.\ (\ref{eq:Ppm}) and $\tau$, $\tau' \in \{ {\rm e},{\rm h}\}$. For $\tilde E = 0$ one has $\cos \bar \eta = 0$, so that the equal-spin Andreev reflection amplitude $r_{{\rm h}+,{\rm e}+}(0) = r_{{\rm e}+,{\rm h}+}(0) = 0$, consistent with the general principle that Andreev reflection must be absent in topologically trivial half-metal--superconductor junctions \cite{Beri2009}.

The conductance $G(V)$ of the junction at bias voltage $V$ is \cite{Blonder1982, Jong1995},
\begin{align}
  G(V) =&\, \frac{e^2}{h} \sum_{\sigma,\sigma' = \uparrow,\downarrow}
  \int_{-\infty}^{\infty} d\tilde E 
  \left( - \frac{d f_{\sigma'}(\tilde E - e V)}{d\tilde E} \right)
  \nonumber \\ &\, \mbox{} \times
  (\delta_{\sigma \sigma'} -
  |r_{{\rm e}\sigma,{\rm e}\sigma'}(\tilde E)|^2 + |r_{{\rm h}\sigma,{\rm e}\sigma'}(\tilde E)|^2),
\end{align}
where the spin labels $\sigma$ and $\sigma'$ refer to the spin projection onto the $z$ axis. Substituting Eq.\ (\ref{eq:r_HM}), taking the limit $V \to 0$, $T \to 0$, where $T$ is temperature, and expanding to leading order in the precession frequency $\Omega$, we find that the conductance $G$ of the half metal--superconductor junction is
\begin{align}
  G(V) =&\, \frac{e^2}{h} \frac{2(\hbar \Omega)^2}{|\Delta|^2} \sin^2 \theta
  \nonumber \\ &\, \mbox{} \times
  \left[(eV)^2 + (\hbar \Omega)^2 + \frac{\pi^2}{3} (k_{\rm B}T)^2 \right] 
\end{align}
in the limit of small $V$, $T$, and $\Omega$. Note that the zero-bias, zero-temperature conductance involves Andreev reflection coefficients at rotating-frame energy $\tilde E = -\tfrac{1}{2} \hbar \Omega$, which is why $G(0)$ is nonzero at $T = 0$ despite the absence of Andreev reflection for $\tilde E = 0$.

The Andreev reflection amplitudes also determine the anomalous Green function \cite{McMillan1968}. In the rotating frame and in the spin-majority/minority basis ($s,s' = \pm$), the retarded anomalous Green function is
\begin{equation}
  \label{eq:F}
    \tilde{F}^{\rm R}_{ss'}(x,t; x', t') = -i \Theta(t-t') \langle \{\hat{\tilde{\psi}}_s(x,t), \hat{\tilde{\psi}}_{s'}(x',t') \} \rangle.
\end{equation}
Its Fourier transform $\tilde{F}^{\rm R}(x,x'; \omega)$ is the electron–hole block of the matrix Green function, 
\begin{equation}
    \tilde {\cal G}(\omega + i 0) = \begin{pmatrix} \tilde G^{\rm R}(x,x';\omega) &
    \tilde F^{\rm R}(x,x';\omega) \\ -\tilde F^{\rm R}(x,x';-\omega)^* &
    -\tilde G^{\rm R}(x,x';-\omega)^* \end{pmatrix},
    \label{eq:Gmatrix}
\end{equation}
which satisfies the Gor'kov equation~\cite{Gorkov1958,Gorkov1959}
\begin{equation}
    (\hbar \omega - \tilde{{\cal H}}_{\rm BdG}) \tilde{\mathcal{G}}(x, x'; \omega) = \mathbb{1} \delta(x-x'),
    \label{eq:Gorkov}
\end{equation}
where $\mathbb{1}$ is the identity matrix in spin and particle-hole space. We focus on the induced superconductivity inside the normal region, {\em i.e.}, on $x, x' < -L/2$. According to the Gor'kov equation (\ref{eq:Gorkov}), the anomalous Green function (\ref{eq:F}) can then be calculated as the electron component with spin $s$ at position $x$ of the solution of the Bogoliubov-de Gennes equation with a source that injects a hole with spin $s'$ at position $x'$. This process requires Andreev reflection, so that we find \cite{McMillan1968, Kupferschmidt2011}
\begin{gather}
    \tilde{F}^{\rm R}_{ss'}(x,x'; \omega) = 
    - \frac{i}{\hbar v_0} e^{-i k_0(x-x')} r_{{\rm e}s,{\rm h}s'}(\hbar \omega).
\end{gather}
The advanced component of the anomalous Green function can be calculated from particle-hole symmetry
\begin{equation}
  \tilde F^{\rm A}_{ss'}(x,x';\omega) = - \tilde F^{\rm R}_{s's}(x',x;-\omega),
  \label{eq:FAR}
\end{equation}
and yields
\begin{equation}
    \tilde{F}^{\rm A}_{ss'}(x,x'; \omega) = 
    \frac{i}{\hbar v_0} e^{i k_0(x-x')} r_{{\rm e}s',{\rm h}s}(-\hbar \omega).
\end{equation}
The expression for the anomalous Green's function can be cast in the form
\begin{equation}
    \tilde{F}_{ss'}(x,x'; \omega) = -\frac{i}{2} \int dk 
    e^{-ik(x-x')} \tilde f_{ss'}(k; \omega) \delta(\varepsilon_k - \omega),
    \label{eq:Fss}
\end{equation}
where $\varepsilon_k = \tfrac{\hbar^2}{2m}(k^2-k_0^2)$ and $\tilde f(k;\omega)$ is the similar to the anomalous Green function in the quasiclassical theory \cite{McMillan1968, Rammer1986},
\begin{align}
  \label{eq:fss}
    \tilde f_{ss'} (k; \omega)= 2 
    \times \begin{cases}
        r_{{\rm e}s, {\rm h}s'}(\omega), & k < 0, \, \Im \omega > 0, \\
        -r_{{\rm e}s', {\rm h}s}(-\omega), & k > 0, \, \Im \omega < 0. \\
        0, & \text{else}
    \end{cases}
\end{align}
Note that $f_{ss'}(k,0) = 0$, because the Andreev reflection amplitudes vanish at $\tilde E = 0$.
Equation (\ref{eq:fss}) combines the results for both the retarded and advanced Green functions.

For the half metal--superconductor junction, only the anomalous Green function component $\tilde f_{++}(k;\omega)$ for $s = s' = +$ is non-zero. This component is even under exchange of the spin labels. To bring about the symmetry of the proximity superconductivity with respect to the momentum $k$ and the frequency $\omega$ we write $\tilde f_{ss'}(k;\omega)$ as a sum of a $k$-odd, $\omega$-even and an $\omega$-odd, $k$-even component,
\begin{equation}
  \tilde f_{ss'}(k;\omega) = \tilde f_{ss'}^{+-}(k;\omega) + \tilde f_{ss'}^{-+}(\omega),
\end{equation}
with
\begin{align}
    \tilde f_{ss'}^{+-}(k,\omega) =&\, \delta_{s,s'} \delta_{s,+} \text{sign}(\Im \omega)\,
    r_{{\rm e}+, {\rm h}+}(\omega\, \text{sign}(\Im \omega)), \nonumber \\
    \tilde f_{ss'}^{-+}(k,\omega) =&\, \delta_{s,s'} \delta_{s,+} \text{sign}(k)\, r_{{\rm e}+, {\rm h}+}(\omega\, \text{sign}(\Im \omega)). 
\end{align}
The $k$-even, $\omega$-odd component is generally robust to disorder scattering and, hence, is responsible for the long-range proximity effect \cite{Bergeret2005}. 

\subsection{S/HM/S junction}

For a junction with two superconducting contacts, we first discuss the bound-state spectrum and its contribution to the Josephson current and include the continuous part of the spectrum afterwards. To further simplify the calculation of the Josephson current, we focus on the short-junction limit $L \ll \xi$, where $\xi = \hbar v_0/|\Delta|$ is the superconductor coherence length.

\begin{figure}[t]
\includegraphics[width=0.45\textwidth]{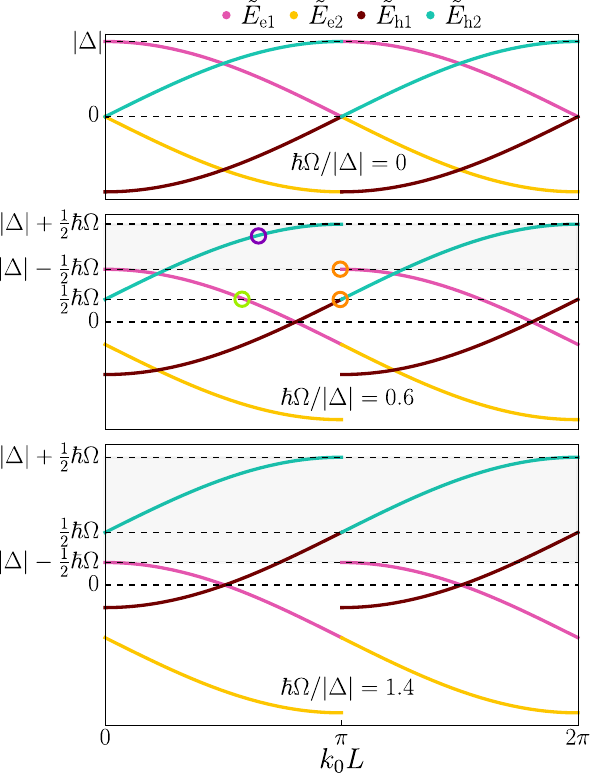}
    \caption{Bound-state energies $\tilde E_{{\rm e}j}$ and $\tilde E_{{\rm h}j}$, $j=1,2$, of a static, i.e. for $\theta = 0$, S/HM/S junction in the short-junction limit $L \ll \xi$ as a function of $k_0L$ for $\hbar\Omega/|\Delta| = 0$ (top), $0.6$ (middle), and $1.4$ (bottom) in the rotating frame. Bound states with energy $\tilde E_\alpha > |\Delta| - \hbar|\Omega|/2$ hybridize with the continuum of scattering states (shaded region). The circles in the middle panel, corresponding to $\hbar\Omega = 0.6|\Delta|$, indicate (i) the example for $k_0 L =2.0$ of a typical point where one of the bound states hybridizes with the continuum spectrum (purple circle), (ii) an abrupt change in the occupation of a bound state (green and bottom orange circles) upon varying $k_0 L$, and (iii) merging of a bound state with the continuum spectrum (top orange circle). For the energy $\tilde E$ corresponding to (i) the energy-resolved current $j_{\uparrow}(\tilde E)$ exhibits a sharp peak, see Fig.~\ref{fig:jE}. Case (ii) leads to a jump in the current as a function of system parameters, as illustrated in the right panel of Fig.~\ref{fig:SHMS}.}
    \label{fig:ABS_HM}
\end{figure}

For $\theta = 0$ one finds electron-like and hole-like bound states at the energy given by the solution(s) of the conditions  $e^{2 i k_0 L - 4 i \eta_+} = 1$ and $e^{-2 i k_0 L - 4 i \eta_-} = 1$, respectively. For more detail, we refer to App.~\ref{app:pert}.
Defining $0 \le \zeta < \pi/2$ such that $e^{4 i \zeta} = e^{2 i k_0 L}$, {\em i.e.},
\begin{equation}
    \zeta = \frac{1}{2} \left(k_0 L \mod \pi \right), \label{eq:HM:zeta}
\end{equation}
in the short-junction limit $L \ll\xi$ we then find two electron-like bound states and the two hole-like bound states at energies
\begin{align} \label{eq:E_e12}
  \tilde E_{{\rm e}1} =&\, - \tilde E_{{\rm h}1} = |\Delta| \cos \zeta - \frac{1}{2}\hbar \Omega, \nonumber \\ \tilde E_{{\rm e}2} =&\, - \tilde E_{{\rm h}2} = {-|\Delta|} \sin \zeta - \frac{1}{2}\hbar \Omega.
\end{align}
Figure \ref{fig:ABS_HM} shows the four bound-state energies $\tilde E_{{\rm e}1}$, $\tilde E_{{\rm e}2}$, $\tilde E_{{\rm h}1}$, and $\tilde E_{{\rm h}2}$ as a function of $k_0 L$ for three representative values of the precession frequency $\Omega$.
The majority wavefunction amplitudes of these bound states are 
\begin{equation}
  \label{eq:atheta0}
  a_{{\rm R}\uparrow j}^{\rm in} = -e^{-2 i \eta_+} a_{{\rm R}\uparrow j}^{\rm out} = \frac{1}{\sqrt{{\cal N}_+}},\ \ b_{{\rm R}\uparrow j}^{\rm out} = b_{{\rm R}\uparrow j}^{\rm in} = 0,
\end{equation}
$j=1,2$, for the electron-like bound states, where the phases $\eta_{\sigma}$ should be evaluated at the respective energies $\tilde E_{{\rm e}j}$ and the normalization constant 
\begin{equation} \label{eq:N_sigma}
  {\cal N}_{\sigma} = 
  \frac{2 \hbar}{|\Delta| \sin \eta_{\sigma}}.
\end{equation}
The minority wavefunction amplitudes follow from Eq.\ (\ref{eq:SS}). The majority and minority wavefunction amplitudes of the hole-like bound states follow by particle-hole conjugation.
The $\theta=0$ bound states exist for arbitrarily short junction length $L$ and have their support entirely in the superconducting regions in the short-junction limit $L \ll \xi$. In this sense they are like standard Andreev bound states, such as they occur in a superconductor--normal metal--superconductor (S/NM/S) junction. 
However, their energies do not depend on the phase difference $\phi$ and, hence, the bound states at $\theta = 0$ do not carry a supercurrent. 

Electron-like bound states exist for $-|\Delta| - \hbar \Omega/2 < |\Delta| - \hbar \Omega/2$; hole-like bound states exist for $-|\Delta| + \hbar \Omega/2 < |\Delta| + \hbar \Omega/2$. For $|\tilde E| < |\Delta| - \hbar |\Omega|/2$ bound states of both types exist. In this energy range, the electron-like bound states acquire a small hole-like component and vice versa for small but non-zero tilt angle $\theta$, whereas the bound state energy $\tilde E$ now depends on the phase difference $\phi$. The hole-like component of the electron-like bound state already appears to linear order in $\theta$, 
\begin{align}
  \label{eq:bRjout}
b_{{\rm R}\uparrow j}^{\rm out} =&\, 
  - \theta B_j 
  \left(e^{-i \eta_-} - e^{i (2\zeta + \eta_- - \phi + \pi j)} \right),
\nonumber \\
b_{{\rm R}\uparrow j}^{\rm in} =&\,  
  \theta B_j
  \left(e^{i \eta_-} - e^{-i (2\zeta + \eta_- + \phi + \pi j)} \right),
\end{align}
where
\begin{equation} \label{eq:B_j}
  B_j =
  -\frac{i}{\sqrt{{\cal N}_+}}
  \frac{\sin \eta' \cos \bar \eta}{\sin(2 (\zeta + \eta_{-}))}
  e^{\tfrac{i}{2} (2\zeta - \pi j)},\ \ j=1,2
\end{equation}
and we omitted corrections beyond linear order in $\theta$.
(The electron-like wavefunctions are still given by Eq.\ (\ref{eq:atheta0}).) We refer to App.~\ref{app:pert} for a more detailed derivation of the wave functions presented above.

The electron-like components of the hole-like bound states again follow by particle-hole conjugation. The $\phi$-dependent contribution to the energies $\delta \tilde E_{{\rm e}j} = -\delta \tilde E_{{\rm h}j}$, $j=1,2$, is quadratic in $\theta$, 
\begin{align}
  \label{eq:dEdphi}
  \frac{\partial \delta \tilde E_{{\rm e}j}}{\partial \phi} = &\,
  2\theta^2 (-1)^{j-1} |\Delta| 
  \frac{\sin \eta_+ (\sin \eta' \cos \bar \eta)^2}{\sin(2(\zeta + \eta_{-}))}
  \sin \phi.
\end{align}
In Eqs.\ (\ref{eq:bRjout})--(\ref{eq:dEdphi}) the phases $\eta_{\sigma}$ should be evaluated at $\tilde E_{{\rm e}1}$ and $\tilde E_{{\rm e}2}$ for $j=1$ and $j=2$, respectively. Note that the factor $\cos^2 \bar \eta$ implies that $\partial \delta \tilde E_{{\rm e}j}/\partial \phi \propto \Omega^2$ for small precession frequency $\Omega$.

Since the Andreev bound states are fully spin-polarized inside the central part of the junction and have electron-like and hole-like parts of equal weight, one has
\begin{equation}
  \rho_{\alpha \uparrow} = \rho_{\alpha \downarrow},
\end{equation}
see Eq.\ (\ref{eq:rhoalpha}). (The label $\alpha$ takes the values ${\rm e}1$, ${\rm e}2$, ${\rm h}1$, and ${\rm h}2$.) The occupation of the bound state $\alpha$ then follows from Eq.\ (\ref{eq:nalpha}),
\begin{equation}
  n_{\alpha} = \frac{1}{2} \left[f_\uparrow(\tilde E_{\alpha}) + f_\downarrow(\tilde E_{\alpha})\right].
  \label{eq:nalphaH}
\end{equation}
At zero temperature $n_{\alpha} = 1/2$ for $|E_{\alpha}| < \hbar \Omega/2$ and the contribution of the bound state $\alpha$ to the current through the junction vanishes, see Eqs.\ (\ref{eq:Itotal}) and (\ref{eq:IEdiscrete}). Since bound states with $\tilde E_{\alpha} > |\Delta| - \hbar |\Omega|/2$ hybridize with the continuum spectrum, as we discuss below, only bound states in the energy range $\hbar |\Omega|/2 < \tilde E_{\alpha} < |\Delta|- \hbar |\Omega|/2$ contribute to the ground-state current with the occupation (\ref{eq:nalphaH}).

\begin{figure}[t]
\includegraphics[width=0.45\textwidth]{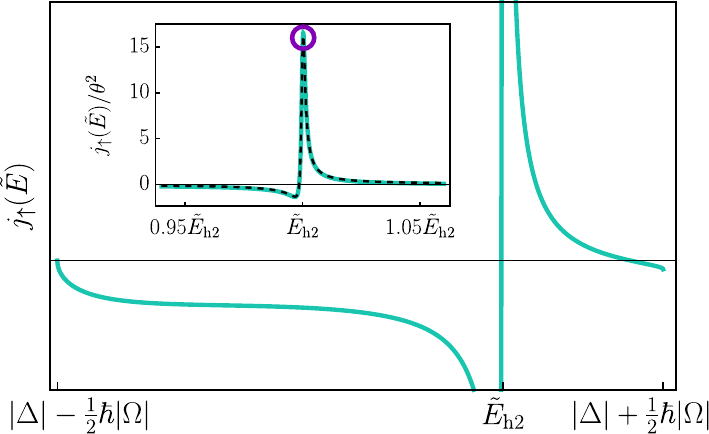}
\caption{Probability current $j_\uparrow(\tilde E)$ carried by the continuum part of the spectrum in the S/HM/S junction ($L \ll \xi$) as a function of energy $\tilde E > |\Delta| - \tfrac{1}{2}\hbar|\Omega|$, for $\hbar\Omega = 0.6|\Delta|$, $\theta = 0.1$, $k_0L = 2.0$, and phase bias $\phi = \pi/2$. The current $j_\uparrow(\tilde E)$ exhibits a sharp peak, with height $\propto \theta^{-2}$ and width $\propto \theta^2$, near the bound-state energy $\tilde E_{{\rm h}2}$ (purple circle; see Fig.~\ref{fig:ABS_HM}). Inset:
A zoom-in near the resonance at $\tilde E_{{\rm h}2}$, together with the lineshape (\ref{eq:Jresonance}) (dashed).
}
    \label{fig:jE}
\end{figure}

For $|\Delta| - \hbar |\Omega|/2 < \tilde E < |\Delta| + \hbar |\Omega|/2$ bound states coexist with scattering states if the tilt angle $\theta = 0$. At finite tilt angle $\theta$ the bound states and scattering states hybridize into scattering resonances, so that for non-zero $\theta$ it is sufficient to consider the scattering states only.
For $\tilde E > |\Delta| + \hbar |\Omega|/2$ there are scattering states of both spin directions $\sigma$, but no bound states and, hence, no scattering resonances.

\begin{figure*}[t]
\centerline{ \includegraphics[width=1.\textwidth]{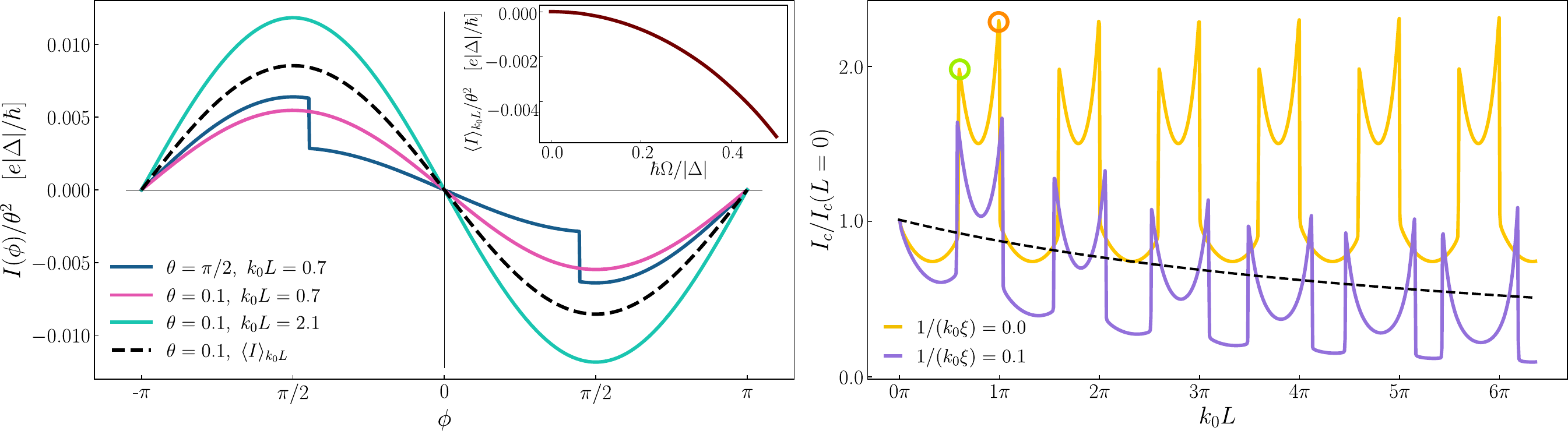}}
\caption{Supercurrent of a one-dimensional S/HM/S Josephson junction at zero temperature. \textit{Left}: current-phase relation in the short-junction limit for different tilt angles and junction lengths, for $\hbar\Omega = 0.6 |\Delta|$. The turquoise and pink curves illustrate the current-phase relation in the limit of small tilt angles for two representative values of $k_0 L$; the dashed curve shows that the average $\langle I \rangle_{k_0 L}$ of $I(\phi)$ over $k_0 L$ at fixed $\phi$ is well-defined and nonzero. The dark blue curve illustrates the large tilt-angle case $\theta = \pi/2$, where a strong deviation from the sinusoidal behavior is observed.
Inset: dependence of $\langle I \rangle_{k_0 L}$ on the precession frequency $\Omega$ at phase bias $\phi = \pi/2$ and $\theta = 0.1$ in the short-junction limit. For small values of $\hbar\Omega/|\Delta|$, the Josephson current has a quadratic dependence on the tilt angle $\theta$ and it vanishes at $\Omega = 0$. \textit{Right}: critical current $I_{\rm c} = \max_{\phi} I(\phi)$ as a function of $k_0L$ for $\hbar\Omega = 0.6 |\Delta|$ and $\theta = 0.1$, shown for the short-junction case $1/(k_0\xi) = 0$ (yellow curve) and in the crossover toward the long-junction limit, $1/(k_0\xi) = 0.1$ (purple curve). The two circles on the yellow curve mark the same points as in Fig.~\ref{fig:ABS_HM}.  They indicate the points where the bound states $\tilde E_{{\rm e}1}$ and $\tilde{E}_{\rm h2}$, respectively, undergo an abrupt change in their occupation, see Eq.~\eqref{eq:nalphaH}. Similar behavior is also seen for $1/(k_0\xi) = 0.1$, although the positions of the jumps are now shifted. In addition, the current is no longer periodic in $k_0L$ and instead shows an overall decaying trend, which is similar to the decay of the supercurrent with $L$ in a superconductor--normal-metal--superconductor junction (dashed line).} 
\label{fig:SHMS}
\end{figure*}

Combining contributions from the discrete and continuous parts of the spectrum,
the Josephson current $I$ through a S/HM/S junction is expressed as
\begin{align}
  \label{eq:IH}
  I =&\, - \frac{e}{\hbar} \sum_{0 \le \tilde E_{\alpha} < |\Delta| - \tfrac{1}{2} \hbar |\Omega|} 
  \frac{\partial \tilde E_{\alpha}}{\partial \phi}
  (1-2n_{\alpha})
  \nonumber \\ &\, \mbox{}
  + \frac{e}{\hbar} 
  \sum_{\sigma} \int_{|\Delta| - \tfrac{1}{2} \hbar \sigma \Omega}^{\infty} d\tilde E j_{\sigma}(\tilde E) (1-2 f_\sigma(\tilde E)),
\end{align}
with $n_{\alpha}$ given by Eq.\ (\ref{eq:nalphaH}). 

For small tilt angle $\theta$, $j_{\sigma}(\tilde E)$ has sharp resonances if $\sigma = \mbox{sign}\, \Omega$ and the energy $|\Delta| - \tfrac{1}{2}\hbar |\Omega| < \tilde E < \Delta| + \tfrac{1}{2}\hbar |\Omega|$ is close to one of the bound state energies $\tilde E_{\alpha}$ at $\theta = 0$. In that case, for small $\theta$ one has
\begin{equation}
  j_{\sigma}(\tilde E) = \theta^2 J_{\alpha}
  \frac{\tilde E - \tilde E_{\alpha} + \theta^2 \delta_{\alpha}}{|\tilde E - \tilde E_{\alpha} + \theta^2 \Gamma_{\alpha}|^2},
  \label{eq:Jresonance}
\end{equation}
where $J_{\alpha}$, $\delta_{\alpha}$, 
and $\Gamma_{\alpha}$ are resonance-specific constants with the dimension of energy. ($J_{\alpha}$ and $\delta_{\alpha}$ are real; $\Gamma_{\alpha}$ is complex.) We refer to App.\ \ref{app:SHMS} for explicit expressions. As an illustration, in Fig.\ \ref{fig:jE} we show $j_{\uparrow}(\tilde E)$ in the short-junction limit for $\hbar \Omega/|\Delta| = 0.6$ and $k_0 L = 2.0$, in which case it has a resonance near the bound state energy $\tilde E_{{\rm h}2}$. (See Fig.\ \ref{fig:ABS_HM} for the bound-state spectrum at $\theta =0$.) We note that Eq.~\eqref{eq:Jresonance} is valid only in the immediate vicinity of the resonance, {\em i.e.}, when $|\tilde E - \tilde E_{\alpha}| \sim \theta^2 |\mbox{Im} \, \Gamma_{\alpha}|$, and requires the resonance width $\theta^2 |\mbox{Im} \, \Gamma_{\alpha}|$ to be much smaller than the other relevant energy scales, such as $\hbar |\Omega|$ or the distance between the resonance position $\tilde E_{\alpha}$ and the continuum edge at $|\Delta| \pm \hbar |\Omega|/2$. When the opposite is true, the full scattering expression should be used instead of the isolated-resonance approximation.

For a small tilt angle $\theta$ the Josephson current we obtain from Eq.\ (\ref{eq:IH}) is proportional to $\theta^2 \sin \phi$. The $\phi$-dependence of the Josephson current is no longer sinusoidal for larger $\theta$, as we illustrate in Fig.\ \ref{fig:SHMS} (left) for a junction in the short-junction limit. A non-sinusoidal current-phase relationship was also observed in Ref.\ \cite{Holmqvist2012} for a Josephson junction coupled to a precessing spin. The jumps in the current-phase relation for large $\theta$ originate from an abrupt change in the steady-state population of an Andreev bound state once its energy crosses the energy splitting $\tfrac{1}{2} \hbar |\Omega|$, see Eq.~\eqref{eq:nalpha} and Eq.~\eqref{eq:nalphaH}. In our analysis, the steady state is established independently for each fixed set of system parameters, such as the superconducting phase difference $\phi$ or the junction length $L$. Therefore, at zero temperature the occupation, and consequently the Josephson current, may change discontinuously as a function of these parameters \footnote{This should be distinguished from a dynamical measurement of the Josephson current, in which the phase difference is swept continuously. In the latter case, the population can remain continuous because the system may stay in a metastable state with a non-equilibrium occupation of the Andreev levels.}. 

The left panel of Fig.\ \ref{fig:SHMS} also shows the average $\langle I(\phi) \rangle_{k_0 L}$ of $I(\phi)$ over $k_0 L$ for the short-junction limit, where $I(\phi)$ is a periodic function of $k_0 L$. The fact that the average $\langle I(\phi)\rangle_{k_0 L}$ is non-zero is an indicator for the long range of the proximity effect in a superconductor--half-metal--superconductor junction with a precessing magnetization. 

In the right panel of Fig.\ \ref{fig:SHMS} we show the critical current $I_{\rm c} = \max_{\phi} I(\phi)$ as a function of $k_0 L$, both in the short-junction limit $L \ll \xi$ and in the crossover towards the long-junction limit. The sharp discontinuities as a function of $k_0 L$ are the result of the abrupt change of population of the Andreev bound states with increasing $k_0 L$ at zero temperature, see Fig.\ \ref{fig:ABS_HM}. The contribution to the Josephson current from the continuous part of the spectrum, which is the second term in Eq.\ (\ref{eq:IH}), is a continuous function of $k_0 L$. Abrupt changes of the Josephson current also appear for the opposite-spin contribution to the Josephson current in junctions with a non-precessing ferromagnet, see Ref.~\cite{Cayssol2004}. 

Our numerical results indicate that the length-averaged current $\langle I \rangle_{k_0 L}$ depends quadratically on the precession frequency $\Omega$, 
\begin{equation}
    \langle I \rangle_{k_0 L} \propto \left(\frac{\hbar \Omega}{|\Delta|}\right)^2 \theta^2 \sin \phi,
\end{equation}
see Fig.~\ref{fig:SHMS}. As will be discussed in Sec.\ \ref{Sec:2D:3D}, the quadratic $\Omega$-depencence of the average $\langle I \rangle_{k_0 L}$ implies a quadratic $\Omega$-dependence of the Josephson current $I$ at fixed $k_0 L$ in two- and three-dimensional junctions. For a one-dimensional junction, we find that the (non-averaged) Josephson current $I$ at a fixed value of $k_0 L$ is proportional to $\Omega^2$ for most values of $k_0 L$, but not for all $k_0 L$. 
An explicit example in which $I(\Omega)$ is not proportional to $\Omega^2$ is shown in Appendix~\ref{app:SHMS:Omega<<Delta}.

\section{N/F/S and S/F/S junctions \label{Sec:SFMS}}

In this Section, we consider a Josephson junction with a partially polarized ferromagnetic metal F and a precessing magnetization. Since a partially polarized magnet admits an opposite-spin proximity effect, such a junction already has a Josephson current without precessing magnetization. In a one-dimensional junction, this opposite-spin contribution to the supercurrent strongly oscillates as a function of the junction length $L$. A precessing magnetization induces spin-polarized proximity superconductivity, which gives a non-oscillating contribution to the Josephson current, which is proportional to $\theta^2$ for small tilt angle $\theta$.

To keep the expressions simple, we take the limit $|\Delta| \ll |h_{\rm F}| \ll \mu$. In this limit, the reflection and transmission matrices of the ferromagnetic metal are
\begin{align} \label{eq:r_t_weakF}
  & r_{\pm}(\tilde E) = 0,\ \
  && t_{\pm}(\tilde E) = e^{i (k_0 \pm \kappa) L + i \tilde E L/\hbar v_0},
\end{align}
where we abbreviated 
\begin{equation}
  \kappa = \frac{h_{\rm F}}{\hbar v_0}.
  \label{eq:kappa}
\end{equation}
We also again focus on the short-junction limit $L \ll \xi$.

\begin{figure}[t]
\includegraphics[width=0.45\textwidth]{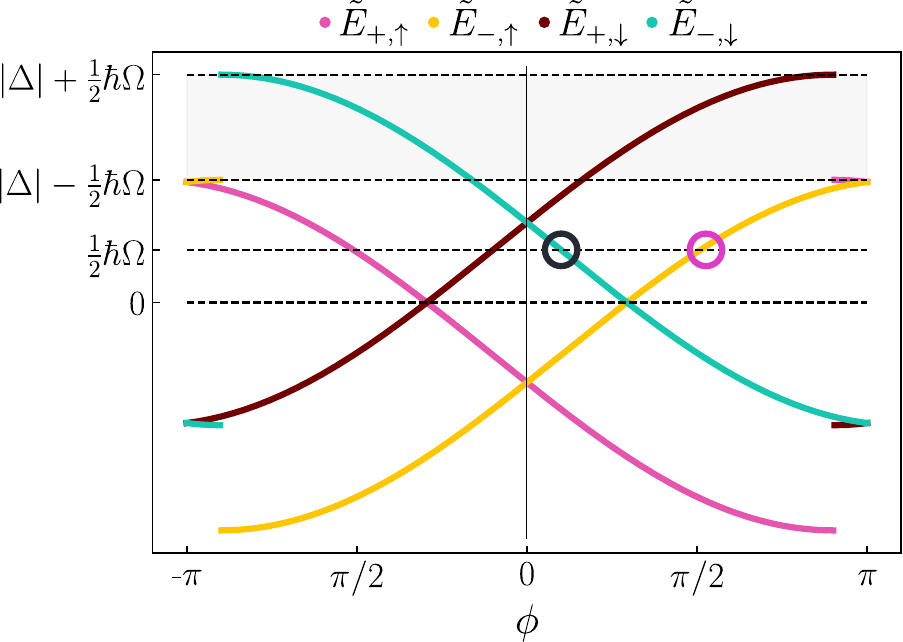}
\caption{Dependence of the bound-state energies $\tilde E_{\pm,\uparrow}$ and $\tilde E_{\pm,\downarrow}$ of a one-dimensional S/F/S Josephson junction on the phase difference $\phi$ for $\hbar\Omega = 0.6|\Delta|$, $\kappa L = 0.55\pi$, and $\theta = 0$.
The black and pink circles mark the points where the occupation $n_{\pm,\sigma}$ of a bound state changes abruptly. Around these points, the current can undergo a sudden change both in the non-precessing case (around the black circle) and in the regime where features distinguishing non-precessing and precessing junctions appear (both black and pink circles), see Fig.~\ref{fig:SFS}.}
    \label{fig:SFS:ABS}
\end{figure}

\subsection{N/F/S junction \label{subsec:SFS:1contact}}

The amplitudes $r_{{\rm h}s,{\rm e}s'}(\tilde E) = r_{{\rm e}s,{\rm h}s'}(-\tilde E)^*$ for Andreev reflection are
\begin{align}\label{eq:r_S}
  r_{{\rm h}+,{\rm e}+}(\tilde E) =&\,
  e^{-i \phi - i \bar \eta} \sin \eta' \sin \theta, \nonumber \\
  r_{{\rm h}-,{\rm e}+}(\tilde E) =&\,
  e^{-i \phi -2 i \kappa L - i \bar \eta}
  (\cos \eta' - i \sin \eta' \cos \theta),\nonumber \\ 
  r_{{\rm h}+,{\rm e}-}(\tilde E) =&\,
  -e^{-i \phi +2 i \kappa L - i \bar \eta}
  (\cos \eta' - i \sin \eta' \cos \theta),\nonumber \\ 
  r_{{\rm h}-,{\rm e}+}(\tilde E) =&\,
  - e^{-i \phi - i \bar \eta} \sin \eta' \sin \theta, 
\end{align}
where we recall that $\bar \eta = (\eta_+ + \eta_-)/2$, $\eta' = (\eta_+ - \eta_-)/2$. The normal reflection amplitudes $r_{{\rm e}s,{\rm e}s'}(\tilde E) = r_{{\rm h}s,{\rm h}s'}(-\tilde E)^*$ all vanish for the simple limit of Eq.~\eqref{eq:r_t_weakF}. Note that the opposite spin-Andreev reflection amplitudes $r_{{\rm h}\mp,{\rm e}\pm}(\tilde E)$ oscillate with the length $L$ of the ferromagnetic segment, whereas the equal-spin Andreev reflection amplitudes $r_{{\rm h}\pm,{\rm e}\pm}(\tilde E)$ do not.

We now use the Andreev reflection amplitudes to calculate the conductance and the anomalous Green function. Since there is no normal reflection, the zero-bias, zero-temperature conductance,
\begin{equation}
    G(0) = 4 \frac{e^2}{h},
\end{equation}
independent of $\theta$ and $\Omega$. This is the same conductance as in an ideal normal-metal--superconductor junction. For the anomalous Green function we again make use of Eqs.\ (\ref{eq:Fss}) and (\ref{eq:fss}). We write the quasiclassical Green function in the rotating frame $\tilde f(k;\omega)$ in matrix form,
\begin{equation}
  \tilde f(k;\omega) = i\sigma_y (\chi(k;\omega) + \bm{d}(k;\omega) \cdot \bm{\sigma}).
\end{equation}
The singlet component $\chi(k;\omega)$ is
\begin{align}
    \chi(k;\omega) =&\,
    \frac{1}{2}
    \sum_{s = \pm} (r_{{\rm e}+,{\rm h}-} (s \omega) - r_{{\rm e}-,{\rm h}+} (s \omega)) \nonumber \\ &\, \ \ \ \ \mbox{} \times 
    \Theta(-s k)\Theta(s \Im \omega).
\end{align}
It contains $k$-even, $\omega$-even and $k$-odd, $\omega$-odd components. The triplet components in the rotating frame read
\begin{align}
    d_x(k;\omega) =&\, -\sum_{s = \pm}s \,r_{{\rm e}+,{\rm h}+} (s \omega)  \Theta(-s k)\Theta(s \Im \omega), \nonumber \\
    d_y(k;\omega) =&\, 0, \\
    d_z(k;\omega) =&\, \frac{1}{2} \sum_{s = \pm} s \,  (r_{{\rm e}+,{\rm h}-} (s \omega) + r_{{\rm e}-,{\rm h}+} (s \omega))
    \nonumber \\ &\, \nonumber \ \ \ \ \mbox{} \times \Theta(-s k)\Theta(s \Im \omega).
\end{align}
They represent non-polarized spin-triplet pairing (given by $d_z$), which vanishes if $h_{\rm F} = 0$ but remains if $\Omega = 0$ or $\theta = 0$, as expected, as well as spin-polarized components:
\begin{gather}
    \tilde f_{++}(k;\omega) = -d_x + i d_y, \quad 
    \tilde f_{--}(k;\omega) = d_x + i d_y.
\end{gather}
Both spin-polarized triplet components vanish without precession. They are sums of $k$-even, $\omega$-odd and $\omega$-even, $k$-odd contributions. Since the opposite-spin Andreev reflection amplitudes oscillate with the length $L$ of the ferromagnetic segment, the spin-singlet and unpolarized triplet contributions to the proximity effect are short-ranged. The equal-spin Andreev reflections do not oscillate with $L$, implying that the spin-polarized proximity superconductivity is long-ranged.

\subsection{S/F/S junction}

\begin{figure*}[t]
\centerline{ \includegraphics[width=1.\textwidth]{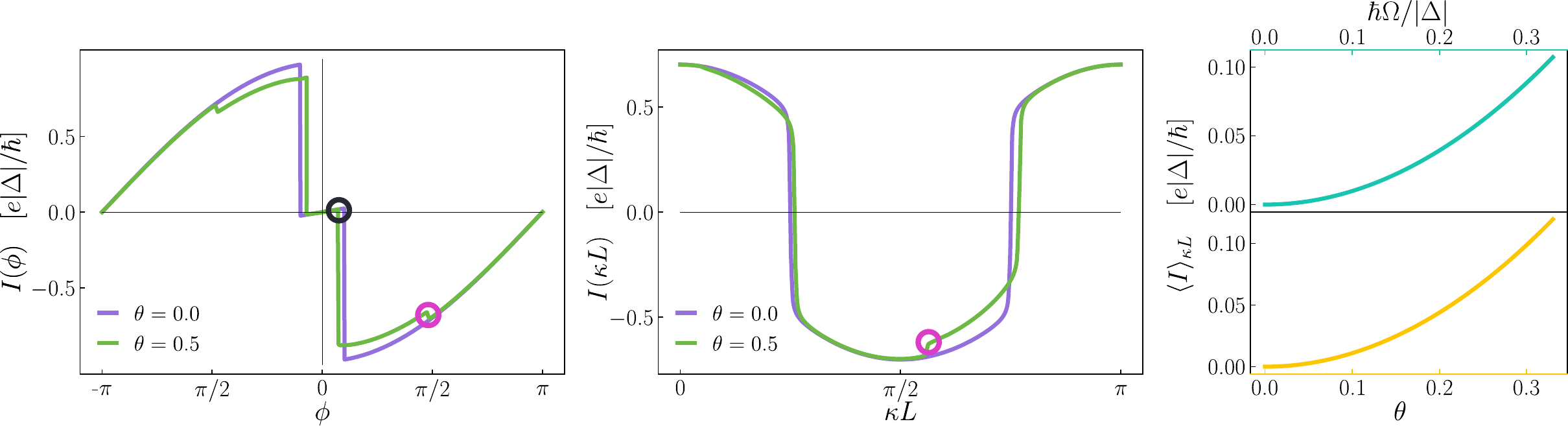}}
\caption{S/F/S Josephson junction at zero temperature in one dimension. \textit{Left}: current-phase relation in the short-junction limit, $1/(\kappa\xi) = 0$, for $\hbar\Omega = 0.6|\Delta|$, $\kappa L = 0.55\pi$, and $\theta = 0$ (violet, non-precessing magnetization) compared to $\theta = 0.5$ (green, precessing magnetization). In the latter case, the shape closely matches the $I(\phi)$ curve of the non-precessing junction, except near the points where the occupation changes abruptly (black and pink circles), cf. Fig.~\ref{fig:SFS:ABS}, where the small-$\theta$ corrections become visible. \textit{Center}: current $I(\kappa L)$ as a function of junction length $\kappa L$ for $\hbar\Omega = 0.6|\Delta|$ at fixed phase bias $\phi = \pi/2$, comparing $\theta = 0$ (violet) and $\theta = 0.5$ (green). The former curve oscillates with $\kappa L$ and averages to zero, implying the short-ranged nature of non-spin-polarized superconductivity in higher dimensions, whereas the precessing junction (green line) is non-zero on average. \textit{Right}: Average $\langle I \rangle_{\kappa L}$ at phase bias $\phi=\pi/2$ in the short-junction limit vs.\ $\theta$ (bottom, $\hbar \Omega/|\Delta| = 0.6$) and $\Omega$ (top, $\theta = 0.5$). 
For small $\hbar\Omega/|\Delta|$ and $\theta$, the current exhibits a quadratic dependence on $\Omega$ and $\theta$, respectively.}
\label{fig:SFS}
\end{figure*}

For $\theta = 0$ one finds Andreev bound states at the energy $\tilde E_{\pm,\sigma}$ given by the solution of $e^{2 i \kappa \sigma L - 2 i \eta_{\sigma} \pm i \phi} = 1$, with $\sigma = 1$ or $\sigma = -1$. (The index $\sigma$ refers to the physical spin of the excitation: The bound state with $\sigma = 1$ has a particle-like part with spin-up and a hole-like part with spin-down; the spins are opposite for the bound state with $\sigma = -1$.) In the short-junction limit, this gives the bound-state energies
\begin{align}
  \tilde E_{\pm,\uparrow} =&\, |\Delta| \cos \zeta_{\pm} - \frac{1}{2} \hbar \Omega, \nonumber \\
  \tilde E_{\pm,\downarrow} =&\, {-|\Delta|} \cos \zeta_{\pm} + \frac{1}{2} \hbar \Omega,
\end{align}
where $0 \le \zeta_{\pm} < \pi$ is defined such that $e^{2 i \zeta_{\pm}} = e^{2 i \kappa L \pm i \phi}$, {\em i.e.},
\begin{equation}
  \zeta_{\pm} = \kappa L \pm \tfrac{1}{2} \phi \mod \pi.
\end{equation}
The Andreev levels $E_{\pm,\sigma}$ are shown in Fig.\ \ref{fig:SFS} (left) for a representative value of $\kappa L$.
The occupation of the bound states is 
\begin{equation}\label{eq:nalphaF}
    n_{\pm,\sigma} = f_{\sigma}(\tilde E_{\pm,\sigma}). 
\end{equation}
At zero temperature, the distribution changes sharply at $\tilde E = \tfrac{1}{2} \hbar |\Omega|$, see Fig.\ \ref{fig:SFS} (left).

Since the bound-state energies $\tilde E_{\pm,\sigma}$ in a partially polarized ferromagnet depend on the phase difference $\phi$ already in the limit $\theta = 0$, the Andreev bound states carry a current,
\begin{align}\label{eq:j_SFS_th=0}
  j_{\pm,\sigma} =&\, - \frac{e}{\hbar}
  \frac{\partial \tilde E_{\pm,\sigma}}{\partial \phi} 
  (1-2 n_{\pm,\sigma}) \nonumber \\
  =&\,
  \pm \frac{e \sigma}{2\hbar}  |\Delta| \sin \zeta_{\pm}
  (1 - 2 f_{\sigma}(\tilde E_{\pm,\sigma}))
  .
\end{align}
Only contributions from levels $\tilde E_{\pm,\sigma} \ge 0$ are taken into account when calculating the supercurrent. 
In addition to the contribution from the discrete Andreev bound states, there is a contribution to the supercurrent from the continuous spectrum for $\tilde E > |\Delta| - \tfrac{1}{2} \hbar |\Omega|$. For the total supercurrent one then arrives at an expression similar to Eq.\ (\ref{eq:IH}), but with
\begin{gather}\label{eq:SFS:continuum}
    j_{\sigma}(\tilde E) = \frac{ -i \sigma \sin(2 \eta_\sigma) \sin (2  \kappa L) \sin \phi}{\pi[\cos(2\eta_\sigma) - \cos(2\zeta_+)][\cos(2 \eta_\sigma)  - \cos(2\zeta_-)]}.
\end{gather}
All contributions to the supercurrent oscillate as a function of $\kappa L$ and the average of the oscillations vanishes. The current-phase relationship is illustrated in Fig.\ \ref{fig:SFS} (center).

\begin{figure*}[t]
\centerline{ \includegraphics[width=1.\textwidth]{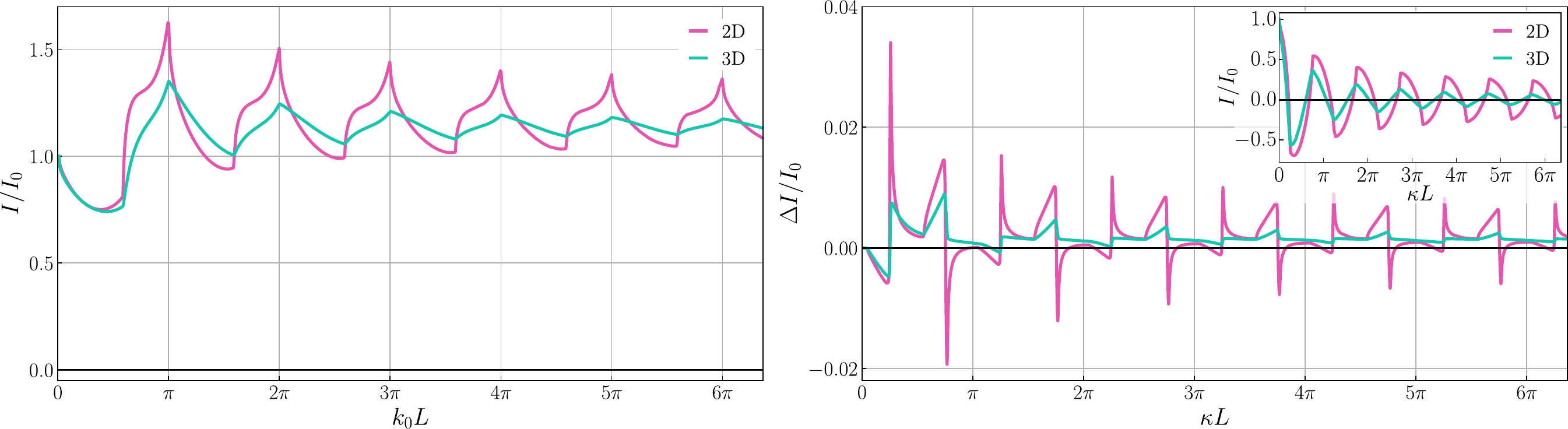}}
\caption{
Josephson current $I(\phi = \pi/2)$ in two- (blue) and three-dimensional (purple) junctions in the short-junction limit $L/\xi \ll 1$ for a fully polarized S/HM/S junction (\textit{left}) and a partially polarized S/F/S junction (\textit{right}). The main right panel shows the difference $\Delta I = I(\Omega) - I(\Omega = 0)$ between the Josephson current with and without precession in F; the inset shows the full current $I$. The parameters used in the calculations are $\hbar\Omega = 0.6|\Delta|$ and $\theta = 0.1$ (left) and $\theta = 0.5$ (right). In the large-$L$ limit, the Josephson current saturates to a nonzero value. For the S/F/S junction, this can be seen from the difference $\Delta I$ in the main panel. In both panels, the current is normalized by its value $I_0$ at $L=0$.
\label{fig:2d_3d}} 
\end{figure*}

A small but non-zero tilt angle $\theta$ leads to a correction $\delta \tilde E_{\pm,\sigma}$ to the bound-state energy that is proportional to $\theta^2$,
\begin{equation}
    \delta \tilde{E}_{\pm,\sigma} =    - \theta^2 \sin \eta_{\sigma} \frac{\sin(\eta_{\sigma} \pm \tfrac{1}{2} \sigma \phi) \sin \eta'}{2 \sin(\bar{\eta} \mp \tfrac{1}{2} \sigma \phi)},
\end{equation}
where $\eta_{\sigma}$, $\eta'$, and $\bar \eta$ should be evaluated at $\tilde E_{\pm,\sigma}$. The additional current carried by the Andreev bound states $\alpha$ is 
\begin{align}
  \delta j_{\pm,\sigma} =&\,
  - \frac{e}{\hbar}
  \frac{\partial \delta \tilde{E}_{\pm,\sigma}}{\partial \phi}
  (1 - 2 f_{\sigma}(\tilde E_{\pm,\sigma}))
  \\ &\,  \nonumber \mbox{}
  -  2 \frac{e}{\hbar}\frac{\partial \tilde{E}_{\pm,\sigma}}{\partial \phi}
  \left( - \frac{\partial f_{\sigma}(\tilde E_{\pm,\sigma})}{\partial \tilde E_{\pm,\sigma}}\right)
  \delta \tilde E_{\pm,\sigma}.
\end{align}
The first term is independent of $\kappa L$ (except for the indirect dependence via the distribution function $f_{\sigma}(\tilde E_{\pm,\sigma})$), so that its average over the length $L$ of the ferromagnetic segment is non-zero. The second term oscillates as a function of $\kappa L$. For small tilt $\theta$ the contribution from the continuous part of the spectrum $\tilde E > |\Delta| - \tfrac{1}{2} \hbar |\Omega|$,  also acquires a change $\delta j_{\sigma}(\tilde E)$, which is proportional to $\theta^2$ and has a non-zero average with respect to $\kappa L$. We refer to the appendix~\ref{app:SFS} for explicit expressions. Figure \ref{fig:SFS} (center) and (left) compare the current-phase relationship and the $\kappa L$-dependence for zero tilt angle and at a finite tilt angle. For small tilt angle $\theta$ the difference is small, except near values of $\phi$ for which the occupation of the Andreev bound states $E_{\pm,\sigma}$ changes. For the average over $\kappa L$ this difference is significant: Whereas the average $\langle I \rangle_{\kappa L}$ vanishes for $\theta = 0$, it is nonzero and proportional to $\theta^2$ for finite $\theta$. For the $\kappa L$-averaged Josephson current, we find the scaling
\begin{equation}
    \langle I_c \rangle_{\kappa L} \propto \left(\frac{\hbar \Omega}{|\Delta|}\right)^2 \theta^2
\end{equation}
for small precession frequencies $\Omega$ and small precession angles $\theta$, see Fig.\ \ref{fig:SFS} (right). The non-vanishing average with respect to $\kappa L$ is key in higher-dimensional junctions, as we discuss in the next Section.

\section{Junctions in two and three dimensions\label{Sec:2D:3D}}

Sofar the discussion has been for one-dimensional junctions. Our results can easily be generalized to two- and three-dimensional junctions. 

For two- and three-dimensional junctions, the momentum components $k_y$ and $k_z$ parallel to the interfaces are conserved and each value of $k_y$ and $k_z$ is separately described by a one-dimensional Hamiltonian of the form of Eq.\ (\ref{eq:H}), but with the substitution 
\begin{equation}
  \mu \to \mu - \frac{\hbar^2}{2 m} k_{\parallel}^2,\ \
  k_{\parallel}^2 = k_y^2 + k_z^2,
  \label{eq:musub}
\end{equation}
in the normal-state first-quantized Hamiltonians of Eqs.\ (\ref{eq:Hnormal}) and (\ref{eq:HF}). (For a two-dimensional junction one sets $k_z = 0$.)
This substitution has no consequences for the boundary conditions of Eqs.\ (\ref{eq:SS}) and (\ref{eq:SScontinuous}) at the superconductor interfaces, as long as the Andreev approximation $|\Delta| \ll \mu$ holds. In the expressions (\ref{eq:r}) and (\ref{eq:t}) for the reflection and transmission amplitudes of the central magnetic segment, one now has to obtain $k$ and $k_{\pm}$ from
\begin{align} \label{eq:kF_ky}
    \frac{\hbar^2 k^2}{2 m} =&\, \mu  + \tilde E - \frac{\hbar^2 k_{\parallel}^2}{2m}, \\ 
    \frac{\hbar^2 k_{\pm}^2}{2 m} =&\, \mu  + \tilde E - V_{\rm F} \pm |h_{\rm eff}| - \frac{\hbar^2 k_{\parallel}^2}{2m}
\end{align}
for each value of $k_y$ and $k_z$ separately. To obtain the total supercurrent in a junction with two superconducting contacts, the sum over all momenta $k_{y,z}$ parallel to the interface must be performed. For a two-dimensional junction this gives
\begin{equation}
  I^{(2{\rm D})} = \int_0^{k_0} \frac{d k_{\parallel}}{\pi} I^{(1{\rm D})}(k_{\parallel}),
  \label{eq:I2d}
\end{equation}
where 
\begin{equation}
    \frac{\hbar^2 k_0^2}{2m} = \mu - V_{\rm F} + |h_{\rm eff}|
\end{equation} and $I^{(1{\rm D})}(k_{\parallel})$ is the current in a one-dimensional junction with the substitution (\ref{eq:musub}). Similarly, the supercurrent in a three-dimensional junction is
\begin{equation}
  I^{(3{\rm D})} = \int_0^{k_0} \frac{k_{\parallel} dk_{\parallel}}{2 \pi} I^{(1{\rm D})}(k_{\parallel}).
  \label{eq:I3d}
\end{equation}
For the calculation of the conductance $G(V)$ and the anomalous Green function of a junction with one normal contact and one superconducting contact, we note that the Andreev reflection amplitudes $r_{\rm eh}$ and $r_{\rm he}$ are now functions of $k_{\parallel}$. The conductance $G(V)$ is found by integration over $k_{\parallel}$ similar to Eqs.\ (\ref{eq:I2d}) and (\ref{eq:I3d}). The anomalous Green function is now a function of $k_x$ and $k_{\parallel}$. It can be found by inserting the $k_{\parallel}$-dependent Andreev reflection amplitudes into Eq.\ (\ref{eq:fss}).

Results for the conductance and the anomalous Green functions in two- and three-dimensional junctions do not differ qualitatively from those for a one-dimensional junction discussed in detail in Secs.\ \ref{subsec:SHMS:1contact} and \ref{subsec:SFS:1contact}. We therefore focus the discussion of the remainder of the Section on the Josephson effect. 

The supercurrent in one-dimensional superconductor--magnet--superconductor junctions shows oscillations as a function of the junction length $L$. For junction involving a half-metal, the oscillations involve the dimensionless product $k_0 L$; for a junction involving a metallic ferromagnet, the oscillations involve the product $\kappa L$, where $\kappa = h_{\rm F}/\hbar v_0 = h_{\rm F} m/\hbar^2 k_0$. In the short-junction limit, the supercurrent is strictly periodic in $k_0 L$ or $\kappa L$ for a junction involving a half metal and a ferromagnetic metal, respectively. Beyond the short junction limit, the supercurrent has a slower dependence on $L/\xi$ in addition to the fast dependence on $k_0 L$ or $\kappa L$. In both cases, since $k_0$ and $\kappa$ depend on $k_{\parallel}$, the integration over the transverse momentum $k_{\parallel}$ in Eqs.\ (\ref{eq:I2d}) and (\ref{eq:I3d}) effectively amounts to an average over $k_0 L$ or $\kappa L$. This is especially true if $k_0 L$ or $\kappa L \gg 1$, indicating that contributions to the supercurrent that vanish upon averaging over $k_0 L$ or $\kappa L$ are short-ranged in a two-dimensional or three-dimensional junction, whereas contributions to the supercurrent that have a nonzero average over $k_0 L$ or $\kappa L$ are long ranged. Our calculations for the one-dimensional junction have shown that the average of the supercurrent $I$ over $k_0 L$ and $\kappa L$ is nonzero if the magnetization precesses. This implies that the Josephson effect is long-ranged in two- and three-dimensional junctions. Without precessing magnetization, the supercurrent is zero in a junction featuring a half metallic magnet. For a junction with a ferromagnetic metal, it vanishes once an average over $\kappa L$ is performed, consistent with the well-known fast decay of the Josephson effect with junction length $L$ in this case.

To illustrate our findings, in Fig.\ \ref{fig:2d_3d} we show the Josephson current $I(\pi/2)$ at phase difference $\phi = \pi/2$ as a function of junction length $L$. We focus on the short-junction limit $L/\xi \ll 1$. Unlike the one-dimensional case, in two and three dimensions, the Josephson current is a non-periodic function of $L$ even in the short-junction limit because of the $k_{\parallel}$-integrations in Eqs.\ (\ref{eq:I2d}) and (\ref{eq:I3d}). The long-range nature of the Josephson current carried by equal-spin Cooper pairs is reflected in the nonzero large-$L$ asymptote within the short-junction limit $L/\xi \ll 1$. Upon going beyond the short-junction limit, a further decay with $L$ sets in, which is similar in nature to that of a non-magnetic Josephson junction \cite{Bergeret2005,Eschrig2015}. For the numerical evaluation, we use the full model as described in Sec.\ \ref{sec:2} and not the simplification used for Eqs.\ (\ref{eq:rHsimpl}) and (\ref{eq:r_t_weakF}), which are used for the numerical results for one-dimensional junctions shown in Secs.\ \ref{Sec:SHMS} and \ref{Sec:SFMS}. (The reason is that these simplifications can not be made simultaneously for all values of $k_{\parallel}$.)

\section{Discussion and conclusion \label{Sec:Discussion}}

In this article we considered the Josephson effect in a ballistic Josephson junction containing a ferromagnetic metal with a precessing uniform magnetization. Without precession, the supercurrent in such a junction is carried by opposite-spin Cooper pairs. It shows decaying oscillations with the junction length $L$, even for a ballistic junction (although for a ballistic junction the suppression is much less than in the presence of disorder): In the short junction limit $L/\xi \ll 1$ one has $I(L) \propto L^{-1/2}$ for a two-dimensional junction and $I(L) \propto L^{-1}$ in three dimensions \cite{Buzdin1982, Buzdin2005}. 
The magnetization precession induces an additional contribution to the supercurrent that is carried by equal-spin Cooper pairs. Although this precession-induced contribution is small for small precession amplitudes --- it is proportional to the square $\theta^2$ of the tilt angle --- it does not suffer from the same suppression with junction length $L$ as the opposite-spin contribution: In the short-junction limit $L \ll \xi$, the precession-induced contribution is independent of $L$. For larger $L$, it decays with $L$ in the same manner as the supercurrent in a superconductor--normal-metal--superconductor junction. 

We could obtain detailed results in the limit of a junction with a half-metallic, {\em i.e.}, fully spin-polarized magnet. In this limit, there is no contribution to the supercurrent from opposite-spin Cooper pairs, so that the magnetization precession switches the junction from an ``off''-state without current to an ``on''-state with finite Andreev conductance and finite Josephson current. For a half metal, the current-phase relationship is sinusoidal for small tilt angle $\theta$, but not for larger tilt angles $\theta \sim 1$. In a partially polarized ferromagnet, the precession-induced term exists next to the supercurrent carried by opposite-spin Cooper pairs. Both have a highly non-sinusoidal current-phase relationship for a short, ballistic Josephson junction \cite{Holmqvist2012}.

Our work can be compared to the original studies of the precession-induced Josephson effect \cite{Takahashi2007,Houzet2008}, which focused on disordered Josephson junctions, and to similar studies for tunneling junctions coupled to a precessing spin \cite{Holmqvist2011,Holmqvist2012,Holmqvist2014}. In particular, our result $I_c \propto \theta^2 \Omega^2$ is reminiscent of Eq.~(14) in Ref.~\cite{Houzet2008}. A key difference, however, is that Refs.~\cite{Takahashi2007,Houzet2008} obtain closed-form expressions for the supercurrent in the vicinity of $T_c$, whereas our analytical results apply to zero temperature. The current-phase relationship is sinusoidal for the diffusive junctions considered in Refs.\ \cite{Takahashi2007,Houzet2008}. For a ballistic junction, we find $I \propto \sin \phi$ only in the limit of small tilt angles $\theta \ll 1$, whereas the current-phase relationship is strongly non-sinuisoidal otherwise. That latter finding is consistent with current-phase relationships obtained in Ref.\ \cite{Holmqvist2012} for a tunneling junction coupled to a precessing spin.

The observation of a long-range proximity effect in the ballistic regime requires that the junction length $L$ be smaller than the mean free path $\ell$, but larger than $\kappa^{-1}$, where $\kappa = (k_{{\rm F}\uparrow} - k_{{\rm F}\downarrow})/2$ is the difference of Fermi wavenumbers for majority and minority electrons, see Eq.\ (\ref{eq:kappa}). For Fe, $\kappa^{-1}$ is approximately $0.3$ nm \cite{Wakoh1966, Schafer2005}, whereas $\ell$ can be up to $10$\,nm in a high-quality thin film~\cite{NguyenTruong2017}. This indeed leaves a parameter window within which ballistic Josephson current can be observed. These lengths are significantly shorter than the superconducting coherence length $\xi$ in an Fe$/$Nb junction, which is $\approx 40$\,nm (using the velocity of minority electrons in Fe $v_0 \approx 3 \cdot 10^5$ m/s~\cite{Schafer2005} and Nb gap $|\Delta| \approx 2.3$meV~\cite{Bonnet1967}), so that a ballistic Nb$/$Fe$/$Nb junction is typically in the short-junction limit. 

As discussed by Takahashi {\em et al.} \cite{Takahashi2007}, in an experimental realization of this scenario, the magnetization precession is typically driven by an alternating applied magnetic field $\vh_{\rm drive}$ at frequency $\Omega$. The tilt angle $\theta$ has a strong resonance near the ferromagnetic resonance frequency $\Omega_{\rm res}$,
\begin{equation} \label{eq:theta_LLG}
    \theta \approx \frac{\gamma h_{\rm drive}/\hbar}{\pi\sqrt{(\Omega - \Omega_{\rm res})^2 + (\alpha \Omega)^2}},
\end{equation}
where $\alpha$ is the Gilbert damping constant of the magnet and $\gamma$ the gyromagnetic ratio. The resonance frequency $\Omega_{\rm res}$ has an intrinsic contribution from the magnetic anisotropy and an additional term $\gamma H_{\rm ext}$, where $\gamma$ is the gyromagnetic ratio and $H_{\rm ext}$ an applied constant magnetic field. Because of the resonant frequency dependence of the tilt angle $\theta$, the supercurrent $I$ (in an S/F/S junction) and the conductance (of an N/F/S junction) have a sharp increase at $\Omega = \Omega_{\rm res}$ upon increasing the driving frequency $\Omega$ at fixed external magnetic field (such that the ferromagnetic resonance frequency $\Omega_{\rm res}$ remains constant). This is illustrated in Fig.\ \ref{fig:G0_res} for the conductance of an N/HM/S junction.

The long-range supercurrents in the presence of a precessing magnetization are spin-polarized. Hence, a finite spin current is associated with the Josephson current. Since the net spin current vanishes in the spin-singlet superconductors, this spin current ``appears'' and ``disappears'' at the superconductor--ferromagnet interfaces. This source or sink of spin current corresponds to a torque on the magnetization. For a steady-state Josephson current, the torques at the two interfaces are equal and opposite, so that their net effect on the total magnetization vanishes.

To optimize the strength of the precession-induced spin-polarized proximity effect, the ratio $\hbar |\Omega|/|\Delta|$ should be as large as possible, but not larger than $2$, to avoid direct excitation of particle-hole pairs in S by the precessing magnetization. With respect to a possible experimental realization, we note that for conventional superconductors, typically $|\Delta|$ is in the range $0.1-3\, \mathrm{meV}$, corresponding to maximum precession frequencies $\Omega_{2 \Delta}/2 \pi$ in the range $0.05-1.5\,\mathrm{THz}$. (For Nb, one has $|\Delta| \approx 2.3\, \mathrm{meV}$~\cite{Bonnet1967}, corresponding to $\Omega_{2\Delta}/2 \pi \approx 1.2\, \mathrm{THz}$, whereas for Al, $|\Delta| \approx 0.3 \mathrm{meV}$~\cite{Biondi1959}, which gives $\Omega_{2 \Delta} \approx 0.1\, \mathrm{THz}$.) For comparison, typical ferromagnetic-resonance frequencies are of the order $\Omega_{\rm res}/2 \pi \sim 1-20\, \mathrm{GHz}$~\cite{Li2025, Golovchanskiy2020, Golovchanskiy2023, Popadiuk2026}, which is small enough to avoid particle-hole excitations, but large enough that the precession-induced contribution is not suppressed by its proportionality to $(\hbar \Omega/|\Delta|)^2$.

The tilt angle $\theta$ is controlled by two main factors: (i) the Gilbert damping parameter $\alpha$, which can be very small (in the half-metals CrO$_2$, Co$_2$MnGe, and Co$_2$MnSi, $\alpha \sim 10^{-4}$~\cite{Liu2009, Zhang2020}, whereas in conventional ferromagnetic metals it is typically $\alpha \sim 10^{-3}$--$10^{-2}$, see, {\em e.g.},~Ref.\ \cite{Chan2023}), and (ii) the ratio $h_{\mathrm{drive}}/H_{\mathrm{ext}}$, which is generally small. As a result, $\theta$ is typically of the order of a few degrees, making it the main small parameter controlling the effect. For observation of the precession-induced effects, half metallic ferromagnets are not only favored because of their low damping, but also because they have no ``conventional'' proximity effect with opposite-spin Cooper pairs, which tends to mask the precession-induced contribution for thin junctions.

Finally, one may wonder about the detrimental effect of the heat dissipated by the precessing magnetization. This could be especially relevant in the case of a ``vertically stacked'' S/F/S junction, because in that case any excess heat in F must be transported away via the superconductors, which are bad heat conductors at low temperatures. Near ferromagnetic resonance, the power dissipated by Gilbert damping (per cross-sectional area of the junction) can be estimated as
\begin{gather} \label{eq:P_Gilbert}
    P_{\rm G} = \frac{\alpha M_s}{\gamma}\Omega_{\rm res}^2 \theta^2 L,
\end{gather}
where $M_s$ is the saturation magnetization (per unit volume).  
Gilbert damping transfers energy from magnons to electrons, typically through spin-orbit coupling or exchange interactions~\cite{Gilmore2007}, thereby heating the conduction electrons in the F region. This energy must be removed through heat transfer to the adjacent superconducting leads, mediated primarily by phonons when $T\ll T_{\rm c}$ and by both phonons and thermally excited quasiparticles when $T\lesssim T_{\rm c}$. At very low temperatures, the bottleneck in heat removal is the transfer of energy from the electronic degrees of freedom to phonons in the F region, since the corresponding thermal conductivity scales as $\propto T^4$ at low temperatures~\cite{Wellstood1994}. On the other hand, when the temperature is close to the critical temperature $T_{\rm c}$, heat can also be transferred directly through thermally excited quasiparticles, without first being transferred to phonons. In either regime, since a sizable long-range proximity effect requires a sizable angle $\theta$, sample heating is an important effect. We expect the effective operating temperature in realistic setups to be comparable to the critical temperatures of conventional superconductors.
 
Several additional effects may significantly influence the driving-induced temperature $\Delta T$ of F at low temperatures. These include the thermal (Kapitza) resistance at the F-S interface, which further impedes heat flow into the superconductor and therefore increases $\Delta T$. Moreover, at lower temperatures, the effective Gilbert damping may be modified by the so-called ``bottleneck effect''~\cite{Hasegawa1959,Barnes1981}, which can decrease the effective value of the damping parameter $\alpha$ in Eq.~\eqref{eq:P_Gilbert}. A more detailed analysis of these contributions, which depend strongly on material-specific properties, is an important step toward the eventual experimental realization of the effects predicted here.

\begin{acknowledgements}
We thank Gal Lemut, Tim Kokkeler, Felix von Oppen, Achim Rosch, and Christoph Strunk for helpful discussions. This work was supported by the Deutsche Forschungsgemeinschaft (DFG, German Research Foundation) as part of the German Excellence Strategy - EXC3112/1 - 533767171 (Center for Chiral Electronics), CRC TR 183 (Project Number 277101999, subproject A02), and the Emmy Noether program - Project Number 506208038.

\end{acknowledgements}

\bibliography{bib-SFS}

@article{Holm1932,
  author  = {R. Holm and W. Meissner},
  title   = {Messungen mit Hilfe von fl{\"u}ssigem Helium. XIII.},
  journal = {Zeitschrift f{\"u}r Physik},
  volume  = {74},
  number  = {11--12},
  pages   = {715--735},
  year    = {1932},
  doi     = {10.1007/BF01340420}
}

@article{Gorkov1958,
  author  = {L. P. Gor'kov},
  title   = {On the Energy Spectrum of Superconductors},
  journal = {Soviet Physics JETP},
  volume  = {7},
  pages   = {505--508},
  year    = {1958}
}

@article{Gorkov1959,
  author  = {L. P. Gor'kov},
  title   = {Microscopic Derivation of the {Ginzburg}-{Landau} Equations in the Theory of Superconductivity},
  journal = {Soviet Physics JETP},
  volume  = {9},
  pages   = {1364--1367},
  year    = {1959}
}

@article{Biondi1959,
  title = {Millimeter Wave Absorption in Superconducting Aluminum. I. Temperature Dependence of the Energy Gap},
  author = {Biondi, Manfred A. and Garfunkel, M. P.},
  journal = {Phys. Rev.},
  volume = {116},
  issue = {4},
  pages = {853--861},
  numpages = {0},
  year = {1959},
  month = {Nov},
  publisher = {American Physical Society},
  doi = {10.1103/PhysRev.116.853},
  url = {https://link.aps.org/doi/10.1103/PhysRev.116.853}
}

@article{Meissner1960,
  title = {Superconductivity of Contacts with Interposed Barriers},
  author = {Meissner, Hans},
  journal = {Phys. Rev.},
  volume = {117},
  issue = {3},
  pages = {672--680},
  numpages = {0},
  year = {1960},
  month = {Feb},
  publisher = {American Physical Society},
  doi = {10.1103/PhysRev.117.672},
  url = {https://link.aps.org/doi/10.1103/PhysRev.117.672}
}

@article{Josephson1962,
title = {Possible new effects in superconductive tunnelling},
journal = {Physics Letters},
volume = {1},
number = {7},
pages = {251-253},
year = {1962},
issn = {0031-9163},
doi = {https://doi.org/10.1016/0031-9163(62)91369-0},
url = {https://www.sciencedirect.com/science/article/pii/0031916362913690},
author = {B.D. Josephson}
}

@article{Werthamer1963,
  title = {Theory of the Superconducting Transition Temperature and Energy Gap Function of Superposed Metal Films},
  author = {Werthamer, N. R.},
  journal = {Phys. Rev.},
  volume = {132},
  issue = {6},
  pages = {2440--2445},
  numpages = {0},
  year = {1963},
  month = {Dec},
  publisher = {American Physical Society},
  doi = {10.1103/PhysRev.132.2440},
  url = {https://link.aps.org/doi/10.1103/PhysRev.132.2440}
}

@article{deGennes1963,
title = {Superconductivity in “normal” metals},
journal = {Physics Letters},
volume = {3},
number = {4},
pages = {168-169},
year = {1963},
issn = {0031-9163},
doi = {https://doi.org/10.1016/0031-9163(63)90401-3},
url = {https://www.sciencedirect.com/science/article/pii/0031916363904013},
author = {de Gennes, P.G. and Guyon, E.}
}

@article{deGennes1964,
  title = {Boundary Effects in Superconductors},
  author = {de Gennes, P. G.},
  journal = {Rev. Mod. Phys.},
  volume = {36},
  issue = {1},
  pages = {225--237},
  numpages = {0},
  year = {1964},
  month = {Jan},
  publisher = {American Physical Society},
  doi = {10.1103/RevModPhys.36.225},
  url = {https://link.aps.org/doi/10.1103/RevModPhys.36.225}
}

@article{Wakoh1966,
  author    = {Wakoh, Shinya and Yamashita, Jiro},
  title     = {Band Structure of Ferromagnetic Iron Self-Consistent Procedure},
  journal   = {Journal of the Physical Society of Japan},
  volume    = {21},
  number    = {9},
  pages     = {1712--1726},
  year      = {1966},
  doi       = {10.1143/JPSJ.21.1712}
}

@article{Pannetier2000,
  author  = {B. Pannetier and H. Courtois},
  title   = {Andreev Reflection and Proximity effect},
  journal = {J. Low Temp. Phys.},
  volume  = {118},
  year    = {2000},
  page    = {599}
}

@article{Andreev1964,
  author  = {A. F. Andreev},
  title   = {The Thermal Conductivity of the Intermediate State in Superconductors},
  journal = {Soviet Physics JETP},
  volume  = {19},
  number  = {6},
  year    = {1964},
  pages   = {1228--1232}
}

@article{Bonnet1967,
    title = {A new measurement of the energy gap in superconducting niobium},
    journal = {Physics Letters A},
    volume = {25},
    number = {6},
    pages = {452-453},
    year = {1967},
    issn = {0375-9601},
    doi = {https://doi.org/10.1016/0375-9601(67)90076-X},
    url = {https://www.sciencedirect.com/science/article/pii/037596016790076X},
    author = {D. Bonnet and S. Erlenkämper and H. Germer and H. Rabenhorst}
}

@article{Deutscher1969,
  author       = {Deutscher, G and de Gennes, P G},
  title        = {PROXIMITY EFFECTS.},
  url          = {https://www.osti.gov/biblio/4842932},
  journal      = {pp 1005-34 of Superconductivity.   Vols. 1 and 2.   Parks, R. D. (ed.).   New York, Marcel Dekker, Inc., 1969.},
  place        = {Country unknown/Code not available},
  year         = {1969},
  month        = {10}}

@article{Kulik1970,
  author  = {O. Kulik},
  title   = {Macroscopic Quantization and the Proximity Effect in {S-N-S} Junctions},
  journal = {Soviet Physics JETP},
  volume  = {30},
  pages   = {944},
  year    = {1970},
  note    = {Russian original published in 1969}
}

@article{Jackel1974,
  title = {Direct measurement of current-phase relations in superconducting weak links},
  author = {Jackel, L. D. and Buhrman, R. A. and Webb, W. W.},
  journal = {Phys. Rev. B},
  volume = {10},
  issue = {7},
  pages = {2782--2785},
  numpages = {0},
  year = {1974},
  month = {Oct},
  publisher = {American Physical Society},
  doi = {10.1103/PhysRevB.10.2782},
  url = {https://link.aps.org/doi/10.1103/PhysRevB.10.2782}
}

@article{Buzdin1982,
  author  = {A. I. Buzdin and L. N. Bulaevskii and S. V. Panyukov},
  title   = {Critical-current oscillations as a function of the exchange field and thickness of the ferromagnetic metal ({F}) in an {S-F-S} {Josephson} junction},
  journal = {JETP Letters},
  volume  = {35},
  number  = {4},
  pages   = {178--180},
  year    = {1982}
}

@article{McMillan1968,
  title = {Theory of Superconductor---Normal-Metal Interfaces},
  author = {McMillan, W. L.},
  journal = {Phys. Rev.},
  volume = {175},
  issue = {2},
  pages = {559--568},
  numpages = {0},
  year = {1968},
  month = {Nov},
  publisher = {American Physical Society},
  doi = {10.1103/PhysRev.175.559},
  url = {https://link.aps.org/doi/10.1103/PhysRev.175.559}
}

@article{Blonder1982,
  title = {Transition from metallic to tunneling regimes in superconducting microconstrictions: Excess current, charge imbalance, and supercurrent conversion},
  author = {Blonder, G. E. and Tinkham, M. and Klapwijk, T. M.},
  journal = {Phys. Rev. B},
  volume = {25},
  issue = {7},
  pages = {4515--4532},
  numpages = {0},
  year = {1982},
  month = {Apr},
  publisher = {American Physical Society},
  doi = {10.1103/PhysRevB.25.4515},
  url = {https://link.aps.org/doi/10.1103/PhysRevB.25.4515}
}

@article{Rammer1986,
  title = {Quantum field-theoretical methods in transport theory of metals},
  author = {Rammer, J. and Smith, H.},
  journal = {Rev. Mod. Phys.},
  volume = {58},
  issue = {2},
  pages = {323--359},
  numpages = {0},
  year = {1986},
  month = {Apr},
  publisher = {American Physical Society},
  doi = {10.1103/RevModPhys.58.323},
  url = {https://link.aps.org/doi/10.1103/RevModPhys.58.323}
}

@Article{beenakker1991d,
  title = {Universal limit of critical-current fluctuations in mesoscopic {Josephson} junctions},
  author = {Beenakker, C. W. J.},
  journal = {Phys. Rev. Lett.},
  volume = {67},
  OPTnumber = {27},
  pages = {3836--3839},
  OPTnumpages = {3},
  year = {1991},
  OPTmonth = {Dec},
  doi = {10.1103/PhysRevLett.67.3836},
  publisher = {American Physical Society}
}

@InProceedings{Beenakker1992,
author= {Beenakker, C. W. J.},
editor= {Fukuyama, Hidetoshi
and Ando, Tsuneya},
title= {Three ``Universal'' Mesoscopic {Josephson} Effects},
booktitle= {Transport Phenomena in Mesoscopic Systems},
year={1992},
publisher="Springer Berlin Heidelberg",
address="Berlin, Heidelberg",
pages="235--253",
isbn="978-3-642-84818-6"
}

@article{Wellstood1994,
  title = {Hot-electron effects in metals},
  author = {Wellstood, F. C. and Urbina, C. and Clarke, John},
  journal = {Phys. Rev. B},
  volume = {49},
  issue = {9},
  pages = {5942--5955},
  numpages = {0},
  year = {1994},
  month = {Mar},
  publisher = {American Physical Society},
  doi = {10.1103/PhysRevB.49.5942},
  url = {https://link.aps.org/doi/10.1103/PhysRevB.49.5942}
}

@article{Jong1995,
  title = {Andreev Reflection in Ferromagnet-Superconductor Junctions},
  author = {de Jong, M. J. M. and Beenakker, C. W. J.},
  journal = {Phys. Rev. Lett.},
  volume = {74},
  issue = {9},
  pages = {1657--1660},
  numpages = {0},
  year = {1995},
  month = {Feb},
  publisher = {American Physical Society},
  doi = {10.1103/PhysRevLett.74.1657},
  url = {https://link.aps.org/doi/10.1103/PhysRevLett.74.1657}
}

@article{Gueron1996,
  title = {Superconducting Proximity Effect Probed on a Mesoscopic Length Scale},
  author = {Gu\'eron, S. and Pothier, H. and Birge, Norman O. and Esteve, D. and Devoret, M. H.},
  journal = {Phys. Rev. Lett.},
  volume = {77},
  issue = {14},
  pages = {3025--3028},
  numpages = {0},
  year = {1996},
  month = {Sep},
  publisher = {American Physical Society},
  doi = {10.1103/PhysRevLett.77.3025},
  url = {https://link.aps.org/doi/10.1103/PhysRevLett.77.3025}
}

@article{Fogelstrom2000,
  title = {Josephson currents through spin-active interfaces},
  author = {Fogelstr\"om, Mikael},
  journal = {Phys. Rev. B},
  volume = {62},
  issue = {17},
  pages = {11812--11819},
  numpages = {0},
  year = {2000},
  month = {Nov},
  publisher = {American Physical Society},
  doi = {10.1103/PhysRevB.62.11812},
  url = {https://link.aps.org/doi/10.1103/PhysRevB.62.11812}
}

@article{Tkachov2001,
  title = {Subgap transport in ferromagnet-superconductor junctions due to magnon-assisted {Andreev} reflection},
  author = {Tkachov, Grygoriy and McCann, Edward and Fal'ko, Vladimir I.},
  journal = {Phys. Rev. B},
  volume = {65},
  issue = {2},
  pages = {024519},
  numpages = {15},
  year = {2001},
  month = {Dec},
  publisher = {American Physical Society},
  doi = {10.1103/PhysRevB.65.024519},
  url = {https://link.aps.org/doi/10.1103/PhysRevB.65.024519}
}

@article{Bergeret2001,
  title = {Long-Range Proximity Effects in Superconductor-Ferromagnet Structures},
  author = {Bergeret, F. S. and Volkov, A. F. and Efetov, K. B.},
  journal = {Phys. Rev. Lett.},
  volume = {86},
  issue = {18},
  pages = {4096--4099},
  numpages = {0},
  year = {2001},
  month = {Apr},
  publisher = {American Physical Society},
  doi = {10.1103/PhysRevLett.86.4096},
  url = {https://link.aps.org/doi/10.1103/PhysRevLett.86.4096}
}

@article{Kadigrobov2001,
  author  = {A. Kadigrobov and R. I. Shekhter and M. Jonson},
  title   = {Quantum spin fluctuations as a source of long-range proximity effects in diffusive ferromagnet-superconductor structures},
  journal = {Europhysics Letters},
  volume  = {54},
  number  = {3},
  pages   = {394--400},
  year    = {2001},
  doi     = {10.1209/epl/i2001-00107-2}
}

@article{Eschrig2003,
  title = {Theory of Half-Metal/Superconductor Heterostructures},
  author = {Eschrig, M. and Kopu, J. and Cuevas, J. C. and Sch\"on, Gerd},
  journal = {Phys. Rev. Lett.},
  volume = {90},
  issue = {13},
  pages = {137003},
  numpages = {4},
  year = {2003},
  month = {Apr},
  publisher = {American Physical Society},
  doi = {10.1103/PhysRevLett.90.137003},
  url = {https://link.aps.org/doi/10.1103/PhysRevLett.90.137003}
}

@article{Cayssol2004,
  title = {Exchange-induced ordinary reflection in a single-channel superconductor-ferromagnet-superconductor junction},
  author = {Cayssol, J\'er\^ome and Montambaux, Gilles},
  journal = {Phys. Rev. B},
  volume = {70},
  issue = {22},
  pages = {224520},
  numpages = {8},
  year = {2004},
  month = {Dec},
  publisher = {American Physical Society},
  doi = {10.1103/PhysRevB.70.224520},
  url = {https://link.aps.org/doi/10.1103/PhysRevB.70.224520}
}

@article{Krawiec2004,
  title = {Current-carrying {Andreev} bound states in a superconductor-ferromagnet proximity system},
  author = {Krawiec, M. and Gy\"orffy, B. L. and Annett, J. F.},
  journal = {Phys. Rev. B},
  volume = {70},
  issue = {13},
  pages = {134519},
  numpages = {5},
  year = {2004},
  month = {Oct},
  publisher = {American Physical Society},
  doi = {10.1103/PhysRevB.70.134519},
  url = {https://link.aps.org/doi/10.1103/PhysRevB.70.134519}
}

@article{Bergeret2005,
  title = {Odd triplet superconductivity and related phenomena in superconductor-ferromagnet structures},
  author = {Bergeret, F. S. and Volkov, A. F. and Efetov, K. B.},
  journal = {Rev. Mod. Phys.},
  volume = {77},
  issue = {4},
  pages = {1321--1373},
  numpages = {0},
  year = {2005},
  month = {Nov},
  publisher = {American Physical Society},
  doi = {10.1103/RevModPhys.77.1321},
  url = {https://link.aps.org/doi/10.1103/RevModPhys.77.1321}
}

@article{Buzdin2005,
  title = {Proximity effects in superconductor-ferromagnet heterostructures},
  author = {Buzdin, A. I.},
  journal = {Rev. Mod. Phys.},
  volume = {77},
  issue = {3},
  pages = {935--976},
  numpages = {0},
  year = {2005},
  month = {Sep},
  publisher = {American Physical Society},
  doi = {10.1103/RevModPhys.77.935},
  url = {https://link.aps.org/doi/10.1103/RevModPhys.77.935}
}

@article{Schafer2005,
  title = {Fermi surface and electron correlation effects of ferromagnetic iron},
  author = {Sch\"afer, J. and Hoinkis, M. and Rotenberg, Eli and Blaha, P. and Claessen, R.},
  journal = {Phys. Rev. B},
  volume = {72},
  issue = {15},
  pages = {155115},
  numpages = {11},
  year = {2005},
  month = {Oct},
  publisher = {American Physical Society},
  doi = {10.1103/PhysRevB.72.155115},
  url = {https://link.aps.org/doi/10.1103/PhysRevB.72.155115}
}

@article{Sosnin2006,
  title = {Superconducting Phase Coherent Electron Transport in Proximity Conical Ferromagnets},
  author = {Sosnin, I. and Cho, H. and Petrashov, V. T. and Volkov, A. F.},
  journal = {Phys. Rev. Lett.},
  volume = {96},
  issue = {15},
  pages = {157002},
  numpages = {4},
  year = {2006},
  month = {Apr},
  publisher = {American Physical Society},
  doi = {10.1103/PhysRevLett.96.157002},
  url = {https://link.aps.org/doi/10.1103/PhysRevLett.96.157002}
}

@article{Keizer2006,
  author  = {Keizer, R. S. and Goennenwein, S. T. B. and Klapwijk, T. M. and Miao, G. and Xiao, G. and Gupta, A.},
  title   = {A spin triplet supercurrent through the half-metallic ferromagnet {CrO$_2$}},
  journal = {Nature},
  year    = {2006},
  volume  = {439},
  pages   = {825--827},
  doi     = {10.1038/nature04499},
  url     = {https://doi.org/10.1038/nature04499},
  note    = {Received 22 July 2005; Accepted 29 November 2005; Issue date 16 February 2006}
}

@article{Braude2007,
  title = {Fully Developed Triplet Proximity Effect},
  author = {Braude, V. and Nazarov, Yu. V.},
  journal = {Phys. Rev. Lett.},
  volume = {98},
  issue = {7},
  pages = {077003},
  numpages = {4},
  year = {2007},
  month = {Feb},
  publisher = {American Physical Society},
  doi = {10.1103/PhysRevLett.98.077003},
  url = {https://link.aps.org/doi/10.1103/PhysRevLett.98.077003}
}

@article{Houzet2007,
  title = {Long range triplet {Josephson} effect through a ferromagnetic trilayer},
  author = {Houzet, M. and Buzdin, A. I.},
  journal = {Phys. Rev. B},
  volume = {76},
  issue = {6},
  pages = {060504},
  numpages = {4},
  year = {2007},
  month = {Aug},
  publisher = {American Physical Society},
  doi = {10.1103/PhysRevB.76.060504},
  url = {https://link.aps.org/doi/10.1103/PhysRevB.76.060504}
}

@article{Gilmore2007,
  title = {Identification of the Dominant Precession-Damping Mechanism in Fe, Co, and Ni by First-Principles Calculations},
  author = {Gilmore, K. and Idzerda, Y. U. and Stiles, M. D.},
  journal = {Phys. Rev. Lett.},
  volume = {99},
  issue = {2},
  pages = {027204},
  numpages = {4},
  year = {2007},
  month = {Jul},
  publisher = {American Physical Society},
  doi = {10.1103/PhysRevLett.99.027204},
  url = {https://link.aps.org/doi/10.1103/PhysRevLett.99.027204}
}

@article{Takahashi2007,
  title = {Supercurrent Pumping in {Josephson} Junctions with a Half-Metallic Ferromagnet},
  author = {Takahashi, S. and Hikino, S. and Mori, M. and Martinek, J. and Maekawa, S.},
  journal = {Phys. Rev. Lett.},
  volume = {99},
  issue = {5},
  pages = {057003},
  numpages = {4},
  year = {2007},
  month = {Aug},
  publisher = {American Physical Society},
  doi = {10.1103/PhysRevLett.99.057003},
  url = {https://link.aps.org/doi/10.1103/PhysRevLett.99.057003}
}

@article{Hikino2008,
author = {Hikino ,Shin-ichi and Mori ,Michiyasu and Takahashi ,Saburo and Maekawa ,Sadamichi},
title = {Ferromagnetic Resonance Induced {Josephson} Current in a Superconductor/Ferromagnet/Superconductor Junction},
journal = {J. Phys. Soc. Jpn.},
volume = {77},
number = {5},
pages = {053707},
year = {2008},
doi = {10.1143/JPSJ.77.053707}
}

@article{Houzet2008,
  title = {Ferromagnetic {Josephson} Junction with Precessing Magnetization},
  author = {Houzet, Manuel},
  journal = {Phys. Rev. Lett.},
  volume = {101},
  issue = {5},
  pages = {057009},
  numpages = {4},
  year = {2008},
  month = {Aug},
  publisher = {American Physical Society},
  doi = {10.1103/PhysRevLett.101.057009},
  url = {https://link.aps.org/doi/10.1103/PhysRevLett.101.057009}
}

@article{Braude2008,
  title = {Triplet {Josephson}   Effect with Magnetic Feedback in a Superconductor-Ferromagnet Heterostructure},
  author = {Braude, V. and Blanter, Ya. M.},
  journal = {Phys. Rev. Lett.},
  volume = {100},
  issue = {20},
  pages = {207001},
  numpages = {4},
  year = {2008},
  month = {May},
  publisher = {American Physical Society},
  doi = {10.1103/PhysRevLett.100.207001},
  url = {https://link.aps.org/doi/10.1103/PhysRevLett.100.207001}
}

@article{Beri2009,
  title = {Quantum limit of the triplet proximity effect in half-metal--superconductor junctions},
  author = {B\'eri, B. and Kupferschmidt, J. N. and Beenakker, C. W. J. and Brouwer, P. W.},
  journal = {Phys. Rev. B},
  volume = {79},
  issue = {2},
  pages = {024517},
  numpages = {11},
  year = {2009},
  month = {Jan},
  publisher = {American Physical Society},
  doi = {10.1103/PhysRevB.79.024517},
  url = {https://link.aps.org/doi/10.1103/PhysRevB.79.024517}
}

@article{Liu2009,
    author = {Liu, Chunsheng and Mewes, Claudia K. A. and Chshiev, Mairbek and Mewes, Tim and Butler, William H.},
    title = {Origin of low {Gilbert} damping in half metals},
    journal = {Applied Physics Letters},
    volume = {95},
    number = {2},
    pages = {022509},
    year = {2009},
    month = {07},
    issn = {0003-6951},
    doi = {10.1063/1.3157267},
    url = {https://doi.org/10.1063/1.3157267}
}

@article{Sau2009,
  title = {Generic New Platform for Topological Quantum Computation Using Semiconductor Heterostructures},
  author = {Sau, Jay D. and Lutchyn, Roman M. and Tewari, Sumanta and Das Sarma, S.},
  journal = {Phys. Rev. Lett.},
  volume = {104},
  issue = {4},
  pages = {040502},
  numpages = {4},
  year = {2010},
  month = {Jan},
  publisher = {American Physical Society},
  doi = {10.1103/PhysRevLett.104.040502},
  url = {https://link.aps.org/doi/10.1103/PhysRevLett.104.040502}
}

@article{Alicea2010,
  title = {Majorana fermions in a tunable semiconductor device},
  author = {Alicea, Jason},
  journal = {Phys. Rev. B},
  volume = {81},
  issue = {12},
  pages = {125318},
  numpages = {10},
  year = {2010},
  month = {Mar},
  publisher = {American Physical Society},
  doi = {10.1103/PhysRevB.81.125318},
  url = {https://link.aps.org/doi/10.1103/PhysRevB.81.125318}
}

@article{Lutchyn2010,
  title = {Majorana Fermions and a Topological Phase Transition in Semiconductor-Superconductor Heterostructures},
  author = {Lutchyn, Roman M. and Sau, Jay D. and Das Sarma, S.},
  journal = {Phys. Rev. Lett.},
  volume = {105},
  issue = {7},
  pages = {077001},
  numpages = {4},
  year = {2010},
  month = {Aug},
  publisher = {American Physical Society},
  doi = {10.1103/PhysRevLett.105.077001},
  url = {https://link.aps.org/doi/10.1103/PhysRevLett.105.077001}
}

@article{Oreg2010,
  title = {Helical Liquids and Majorana Bound States in Quantum Wires},
  author = {Oreg, Yuval and Refael, Gil and von Oppen, Felix},
  journal = {Phys. Rev. Lett.},
  volume = {105},
  issue = {17},
  pages = {177002},
  numpages = {4},
  year = {2010},
  month = {Oct},
  publisher = {American Physical Society},
  doi = {10.1103/PhysRevLett.105.177002},
  url = {https://link.aps.org/doi/10.1103/PhysRevLett.105.177002}
}

@article{Anwar2010,
  title = {Long-range supercurrents through half-metallic ferromagnetic {${\text{CrO}}_{2}$}},
  author = {Anwar, M. S. and Czeschka, F. and Hesselberth, M. and Porcu, M. and Aarts, J.},
  journal = {Phys. Rev. B},
  volume = {82},
  issue = {10},
  pages = {100501},
  numpages = {4},
  year = {2010},
  month = {Sep},
  publisher = {American Physical Society},
  doi = {10.1103/PhysRevB.82.100501},
  url = {https://link.aps.org/doi/10.1103/PhysRevB.82.100501}
}

@article{Khaire2010,
  title = {Observation of Spin-Triplet Superconductivity in {Co}-Based {Josephson} Junctions},
  author = {Khaire, Trupti S. and Khasawneh, Mazin A. and Pratt, W. P. and Birge, Norman O.},
  journal = {Phys. Rev. Lett.},
  volume = {104},
  issue = {13},
  pages = {137002},
  numpages = {4},
  year = {2010},
  month = {Mar},
  publisher = {American Physical Society},
  doi = {10.1103/PhysRevLett.104.137002},
  url = {https://link.aps.org/doi/10.1103/PhysRevLett.104.137002}
}

@article{Wang2010,
  author  = {Jian Wang and Mukesh Singh and Mingliang Tian and et al.},
  title   = {Interplay between superconductivity and ferromagnetism in crystalline nanowires},
  journal = {Nature Physics},
  volume  = {6},
  pages   = {389--394},
  year    = {2010},
  doi     = {10.1038/nphys1621}
}

@article{Robinson2010,
author = {J. W. A. Robinson  and J. D. S. Witt  and M. G. Blamire },
title = {Controlled Injection of Spin-Triplet Supercurrents into a Strong Ferromagnet},
journal = {Science},
volume = {329},
number = {5987},
pages = {59-61},
year = {2010},
doi = {10.1126/science.1189246},
URL = {https://www.science.org/doi/abs/10.1126/science.1189246}}

@article{Sprungmann2010,
  title = {Evidence for triplet superconductivity in {Josephson} junctions with barriers of the ferromagnetic {Heusler} alloy {$\text{Cu}_{2}\text{MnAl}$}},
  author = {Sprungmann, D. and Westerholt, K. and Zabel, H. and Weides, M. and Kohlstedt, H.},
  journal = {Phys. Rev. B},
  volume = {82},
  issue = {6},
  pages = {060505},
  numpages = {4},
  year = {2010},
  month = {Aug},
  publisher = {American Physical Society},
  doi = {10.1103/PhysRevB.82.060505},
  url = {https://link.aps.org/doi/10.1103/PhysRevB.82.060505}
}

@article{Potter2010,
  title = {Multichannel Generalization of {Kitaev's} Majorana End States and a Practical Route to Realize Them in Thin Films},
  author = {Potter, Andrew C. and Lee, Patrick A.},
  journal = {Phys. Rev. Lett.},
  volume = {105},
  issue = {22},
  pages = {227003},
  numpages = {4},
  year = {2010},
  month = {Nov},
  publisher = {American Physical Society},
  doi = {10.1103/PhysRevLett.105.227003},
  url = {https://link.aps.org/doi/10.1103/PhysRevLett.105.227003}
}

@article{Volkov2010,
  title = {Odd spin-triplet superconductivity in a multilayered superconductor-ferromagnet {Josephson} junction},
  author = {Volkov, A. F. and Efetov, K. B.},
  journal = {Phys. Rev. B},
  volume = {81},
  issue = {14},
  pages = {144522},
  numpages = {13},
  year = {2010},
  month = {Apr},
  publisher = {American Physical Society},
  doi = {10.1103/PhysRevB.81.144522},
  url = {https://link.aps.org/doi/10.1103/PhysRevB.81.144522}
}

@article{Duckheim2011,
  title = {Andreev reflection from noncentrosymmetric superconductors and Majorana bound-state generation in half-metallic ferromagnets},
  author = {Duckheim, Mathias and Brouwer, Piet W.},
  journal = {Phys. Rev. B},
  volume = {83},
  issue = {5},
  pages = {054513},
  numpages = {13},
  year = {2011},
  month = {Feb},
  publisher = {American Physical Society},
  doi = {10.1103/PhysRevB.83.054513},
  url = {https://link.aps.org/doi/10.1103/PhysRevB.83.054513}
}

@article{Qi2011,
  title = {Topological insulators and superconductors},
  author = {Qi, Xiao-Liang and Zhang, Shou-Cheng},
  journal = {Rev. Mod. Phys.},
  volume = {83},
  issue = {4},
  pages = {1057--1110},
  numpages = {0},
  year = {2011},
  month = {Oct},
  publisher = {American Physical Society},
  doi = {10.1103/RevModPhys.83.1057},
  url = {https://link.aps.org/doi/10.1103/RevModPhys.83.1057}
}

@article{Xu2011,
  title = {Chern Semimetal and the Quantized Anomalous Hall Effect in {${\mathrm{HgCr}}_{2}{\mathrm{Se}}_{4}$}},
  author = {Xu, Gang and Weng, Hongming and Wang, Zhijun and Dai, Xi and Fang, Zhong},
  journal = {Phys. Rev. Lett.},
  volume = {107},
  issue = {18},
  pages = {186806},
  numpages = {5},
  year = {2011},
  month = {Oct},
  publisher = {American Physical Society},
  doi = {10.1103/PhysRevLett.107.186806},
  url = {https://link.aps.org/doi/10.1103/PhysRevLett.107.186806}
}

@article{Hikino2011,
  author  = {S. Hikino and M. Mori and S. Takahashi and S. Maekawa},
  title   = {Microwave-induced supercurrent in a ferromagnetic {Josephson} junction},
  journal = {Supercond. Sci. Technol.},
  volume  = {24},
  number  = {2},
  pages   = {024008},
  year    = {2011},
  doi     = {10.1088/0953-2048/24/2/024008}
}

@article{Teber2010,
  title = {Transport and magnetization dynamics in a superconductor/single-molecule magnet/superconductor junction},
  author = {Teber, S. and Holmqvist, C. and Fogelstr\"om, M.},
  journal = {Phys. Rev. B},
  volume = {81},
  issue = {17},
  pages = {174503},
  numpages = {18},
  year = {2010},
  month = {May},
  publisher = {American Physical Society},
  doi = {10.1103/PhysRevB.81.174503},
  url = {https://link.aps.org/doi/10.1103/PhysRevB.81.174503}
}

@article{Holmqvist2011,
  title = {Nonequilibrium effects in a {Josephson} junction coupled to a precessing spin},
  author = {Holmqvist, C. and Teber, S. and Fogelstr\"om, M.},
  journal = {Phys. Rev. B},
  volume = {83},
  issue = {10},
  pages = {104521},
  numpages = {18},
  year = {2011},
  month = {Mar},
  publisher = {American Physical Society},
  doi = {10.1103/PhysRevB.83.104521},
  url = {https://link.aps.org/doi/10.1103/PhysRevB.83.104521}
}

@article{Kupferschmidt2011,
  title = {Andreev reflection at half-metal/superconductor interfaces with nonuniform magnetization},
  author = {Kupferschmidt, Joern N. and Brouwer, Piet W.},
  journal = {Phys. Rev. B},
  volume = {83},
  issue = {1},
  pages = {014512},
  numpages = {14},
  year = {2011},
  month = {Jan},
  publisher = {American Physical Society},
  doi = {10.1103/PhysRevB.83.014512},
  url = {https://link.aps.org/doi/10.1103/PhysRevB.83.014512}
}

@article{Anwar2012,
  author  = {M. S. Anwar and M. Veldhorst and A. Brinkman and J. Aarts},
  title   = {Long range supercurrents in ferromagnetic {CrO$_2$} using a multilayer contact structure},
  journal = {Applied Physics Letters},
  volume  = {100},
  pages   = {052602},
  year    = {2012},
  doi     = {10.1063/1.3681138}
}

@article{Hubler2012,
  title = {Observation of {Andreev} Bound States at Spin-active Interfaces},
  author = {H\"ubler, F. and Wolf, M. J. and Scherer, T. and Wang, D. and Beckmann, D. and v. L\"ohneysen, H.},
  journal = {Phys. Rev. Lett.},
  volume = {109},
  issue = {8},
  pages = {087004},
  numpages = {5},
  year = {2012},
  month = {Aug},
  publisher = {American Physical Society},
  doi = {10.1103/PhysRevLett.109.087004},
  url = {https://link.aps.org/doi/10.1103/PhysRevLett.109.087004}
}

@article{Gingrich2012,
  title = {Spin-triplet supercurrent in {Co/Ni} multilayer {Josephson} junctions with perpendicular anisotropy},
  author = {Gingrich, E. C. and Quarterman, P. and Wang, Yixing and Loloee, R. and Pratt, W. P. and Birge, Norman O.},
  journal = {Phys. Rev. B},
  volume = {86},
  issue = {22},
  pages = {224506},
  numpages = {4},
  year = {2012},
  month = {Dec},
  publisher = {American Physical Society},
  doi = {10.1103/PhysRevB.86.224506},
  url = {https://link.aps.org/doi/10.1103/PhysRevB.86.224506}
}

@article{Holmqvist2012,
  title = {Spin-precession-assisted supercurrent in a superconducting quantum point contact coupled to a single-molecule magnet},
  author = {Holmqvist, C. and Belzig, W. and Fogelstr\"om, M.},
  journal = {Phys. Rev. B},
  volume = {86},
  issue = {5},
  pages = {054519},
  numpages = {8},
  year = {2012},
  month = {Aug},
  publisher = {American Physical Society},
  doi = {10.1103/PhysRevB.86.054519},
  url = {https://link.aps.org/doi/10.1103/PhysRevB.86.054519}
}

@article{Banerjee2014,
  author  = {N. Banerjee and J. W. A. Robinson and M. G. Blamire},
  title   = {Reversible control of spin-polarized supercurrents in ferromagnetic {Josephson} junctions},
  journal = {Nature Communications},
  volume  = {5},
  pages   = {4771},
  year    = {2014},
  doi     = {10.1038/ncomms5771}
}

@article{Holmqvist2014,
  title = {Spin-polarized {Shapiro} steps and spin-precession-assisted multiple {Andreev} reflection},
  author = {Holmqvist, C. and Fogelstr\"om, M. and Belzig, W.},
  journal = {Phys. Rev. B},
  volume = {90},
  issue = {1},
  pages = {014516},
  numpages = {9},
  year = {2014},
  month = {Jul},
  publisher = {American Physical Society},
  doi = {10.1103/PhysRevB.90.014516},
  url = {https://link.aps.org/doi/10.1103/PhysRevB.90.014516}
}

@article{Eschrig2015,
  author  = {Matthias Eschrig},
  title   = {Spin-polarized supercurrents for spintronics: a review of current progress},
  journal = {Reports on Progress in Physics},
  volume  = {78},
  number  = {10},
  pages   = {104501},
  year    = {2015},
  doi     = {10.1088/0034-4885/78/10/104501}
}

@article{Linder2015,
  author  = {Jacob Linder and Jason W. A. Robinson},
  title   = {Superconducting spintronics},
  journal = {Nature Physics},
  volume  = {11},
  pages   = {307--315},
  year    = {2015},
  doi     = {10.1038/nphys3242}
}

@article{NguyenTruong2017,
  author  = {Nguyen-Truong, Hieu T.},
  title   = {Electron inelastic mean free path at energies below 100 eV},
  journal = {Journal of Physics: Condensed Matter},
  year    = {2017},
  volume  = {29},
  number  = {21},
  pages   = {215501},
  doi     = {10.1088/1361-648X/aa6b9d},
  url     = {https://doi.org/10.1088/1361-648X/aa6b9d},
  note    = {Published 24 April 2017}
}

@article{Jeon2018,
  author  = {K. R. Jeon and C. Ciccarelli and A. J. Ferguson and et al.},
  title   = {Enhanced spin pumping into superconductors provides evidence for superconducting pure spin currents},
  journal = {Nature Materials},
  volume  = {17},
  pages   = {499--503},
  year    = {2018},
  doi     = {10.1038/s41563-018-0058-9}
}

@article{Li2018,
title = {Possible Evidence for Spin-Transfer Torque Induced by Spin-Triplet Supercurrents},
author = {Lai-Lai Li and Yue-Lei Zhao and Xi-Xiang Zhang and Young Sun},
journal = {Chin. Phys. Lett.},
volume = {35},
number = {7},
pages = {077401-077401},
year = {2018},
issn = {},
doi = {10.1088/0256-307X/35/7/077401}
}

@article{Zhang2020,
  title = {Ultralow Gilbert damping in {{CrO}${}_2$} epitaxial films},
  author = {Zhang, Zhenhua and Cheng, Ming and Yu, Ziyang and Zou, Zhaorui and Liu, Yong and Shi, Jing and Lu, Zhihong and Xiong, Rui},
  journal = {Phys. Rev. B},
  volume = {102},
  issue = {1},
  pages = {014454},
  numpages = {9},
  year = {2020},
  month = {Jul},
  publisher = {American Physical Society},
  doi = {10.1103/PhysRevB.102.014454},
  url = {https://link.aps.org/doi/10.1103/PhysRevB.102.014454}
}

@article{Golovchanskiy2020,
  title = {Magnetization Dynamics in Proximity-Coupled Superconductor-Ferromagnet-Superconductor Multilayers},
  author = {Golovchanskiy, I.A. and Abramov, N.N. and Stolyarov, V.S. and Chichkov, V.I. and Silaev, M. and Shchetinin, I.V. and Golubov, A.A. and Ryazanov, V.V. and Ustinov, A.V. and Kupriyanov, M.Yu.},
  journal = {Phys. Rev. Appl.},
  volume = {14},
  issue = {2},
  pages = {024086},
  numpages = {11},
  year = {2020},
  month = {Aug},
  publisher = {American Physical Society},
  doi = {10.1103/PhysRevApplied.14.024086},
  url = {https://link.aps.org/doi/10.1103/PhysRevApplied.14.024086}
}

@article{Jeon2021,
  author  = {K. R. Jeon and B. K. Hazra and K. Cho and et al.},
  title   = {Long-range supercurrents through a chiral non-collinear antiferromagnet in lateral {Josephson} junctions},
  journal = {Nature Materials},
  volume  = {20},
  pages   = {1358--1363},
  year    = {2021},
  doi     = {10.1038/s41563-021-01061-9}
}

@article{Yang2021,
  author  = {Guang Yang and Chiara Ciccarelli and Jason W. A. Robinson},
  title   = {Boosting spintronics with superconductivity},
  journal = {APL Materials},
  volume  = {9},
  pages   = {050703},
  year    = {2021},
  doi     = {10.1063/5.0048904}
}

@article{Golovchanskiy2023,
  title = {Magnetization Dynamics in Proximity-Coupled Superconductor-Ferromagnet-Superconductor Multilayers. II. Thickness Dependence of the Superconducting Torque},
  author = {Golovchanskiy, I.A. and Abramov, N.N. and Emelyanova, O.V. and Shchetinin, I.V. and Ryazanov, V.V. and Golubov, A.A. and Stolyarov, V.S.},
  journal = {Phys. Rev. Appl.},
  volume = {19},
  issue = {3},
  pages = {034025},
  numpages = {10},
  year = {2023},
  month = {Mar},
  publisher = {American Physical Society},
  doi = {10.1103/PhysRevApplied.19.034025},
  url = {https://link.aps.org/doi/10.1103/PhysRevApplied.19.034025}
}

@article{Jeon2023,
  author  = {K. R. Jeon and B. K. Hazra and J. K. Kim and et al.},
  title   = {Chiral antiferromagnetic {Josephson} junctions as spin-triplet supercurrent spin valves and d.c. {SQUIDs}},
  journal = {Nature Nanotechnology},
  volume  = {18},
  pages   = {747--753},
  year    = {2023},
  doi     = {10.1038/s41565-023-01336-z}
}

@article{Chan2023,
  author  = {Chan, A. K. and Cubukcu, M. and Montiel, X. and others},
  title   = {Controlling spin pumping into superconducting {Nb} by proximity-induced spin-triplet {Cooper} pairs},
  journal = {Communications Physics},
  year    = {2023},
  volume  = {6},
  pages   = {287},
  doi     = {10.1038/s42005-023-01384-w},
  url     = {https://doi.org/10.1038/s42005-023-01384-w}
}

@article{Kulikov2024,
  title = {Resonance phenomena in a nanomagnet coupled to a {Josephson} junction under external periodic drive},
  author = {Kulikov, K. V. and Anghel, D. V. and Nashaat, M. and Dolineanu, M. and Sameh, M. and Shukrinov, Yu. M.},
  journal = {Phys. Rev. B},
  volume = {109},
  issue = {1},
  pages = {014429},
  numpages = {9},
  year = {2024},
  month = {Jan},
  publisher = {American Physical Society},
  doi = {10.1103/PhysRevB.109.014429},
  url = {https://link.aps.org/doi/10.1103/PhysRevB.109.014429}
}

@article{Li2025,
  author  = {Li, Jiaxu and Zhang, Jie and Yang, Guang and Zhao, Weisheng},
  title   = {Efficient shift of ferromagnetic resonance by superconductor gating},
  journal = {Applied Physics Reviews},
  year    = {2025},
  volume  = {12},
  number  = {1},
  pages   = {011419},
  doi     = {10.1063/5.0231497},
  url     = {https://doi.org/10.1063/5.0231497}
}

@Article{buettiker1986,
  author = 	 {M. B\"uttiker},
  title = 	 {Role of quantum coherence in series resistors},
  journal = 	 {Phys. Rev. B},
  year = 	 {1986},
  OPTkey = 	 {},
  volume = 	 {33},
  OPTnumber = 	 {},
  pages = 	 {3020},
  OPTmonth = 	 {},
  OPTnote = 	 {},
  OPTannote = 	 {}
}

@article{Chaou2025,
  author        = {Adam Yanis Chaou and Gal Lemut and Felix von Oppen and Piet W. Brouwer},
  title         = {Proximity superconductivity in chiral kagome antiferromagnets},
  journal       = {arXiv preprint arXiv:2508.08372},
  year          = {2025},
  doi           = {10.48550/arXiv.2508.08372},
  archivePrefix = {arXiv}
}

@software{andriyakhina2026,
  author       = {Andriyakhina, Elizaveta S. and Mansouri, Miad and Breitkreiz, Maxim and Brouwer, Piet W.},
  title        = {Precessing {S/F/S} {Josephson} Current Solver},
  month        = apr,
  year         = 2026,
  publisher    = {Zenodo},
  doi          = {10.5281/zenodo.19551958},
  url          = {https://doi.org/10.5281/zenodo.19551958},
}

@article{Popadiuk2026,
  title = {Mechanism of ferromagnetic resonance in ferromagnet-superconductor trilayers},
  author = {Popadiuk, Dariia and Kharlan, Julia and Kravets, Anatolii and Korenivski, Vladislav and K\l{}os, Jaros\l{}aw W. and Golub, Vladimir},
  journal = {Phys. Rev. B},
  volume = {113},
  issue = {1},
  pages = {014425},
  numpages = {9},
  year = {2026},
  month = {Jan},
  publisher = {American Physical Society},
  doi = {10.1103/blxk-m6pz},
  url = {https://link.aps.org/doi/10.1103/blxk-m6pz}
}

@article{Hasegawa1959,
    author = {Hasegawa, Hiroshi},
    title = {Dynamical Properties of s-d Interaction},
    journal = {Progress of Theoretical Physics},
    volume = {21},
    number = {4},
    pages = {483-500},
    year = {1959},
    month = {04},
    issn = {0033-068X},
    doi = {10.1143/PTP.21.483},
    url = {https://doi.org/10.1143/PTP.21.483},
    eprint = {https://academic.oup.com/ptp/article-pdf/21/4/483/5473523/21-4-483.pdf},
}

@article{Barnes1981,
author = {S.E. Barnes},
title = {Theory of electron spin resonance of magnetic ions in metals},
journal = {Advances in Physics},
volume = {30},
number = {6},
pages = {801--938},
year = {1981},
publisher = {Taylor \& Francis},
doi = {10.1080/00018738100101447},
URL = {https://doi.org/10.1080/00018738100101447},
eprint = {https://doi.org/10.1080/00018738100101447}
}

\newpage
\phantom{}
\newpage
\begin{widetext}
\renewcommand{\appendixname}{}

\appendix

\renewcommand{\thefigure}{S\arabic{figure}}

\appendix


\section{Small-$\theta$ solution for the S/HM/S bound states \label{app:pert}}

Equation~\eqref{eq:bRjout} considers the predominantly electron-like state. We use Eq.~\eqref{eq:SS} to eliminate all minority amplitudes from Eqs.~\eqref{eq:SF} and~\eqref{eq:SFh}. Then, the scattering amplitudes satisfy 
\begin{equation}
  \label{eq:aS}
    \begin{pmatrix} a^{\rm out}_{{\rm L}\uparrow} \\
    - e^{i \eta_{-} - i \phi/2} b^{\rm in}_{{\rm L}\uparrow} \\
    a^{\rm out}_{{\rm R}\uparrow} \\
    - e^{i \eta_{-} + i \phi/2} b^{\rm in}_{{\rm R}\uparrow}
    \end{pmatrix} =\, S_{\rm F}
    \begin{pmatrix} a^{\rm in}_{{\rm L}\uparrow} \\
    - e^{-i \eta_{-} - i \phi/2} b^{\rm out}_{{\rm L}\uparrow} \\
    a^{\rm in}_{{\rm R}\uparrow} \\
    - e^{-i \eta_{-} + i \phi/2} b^{\rm out}_{{\rm R}\uparrow}
    \end{pmatrix}, \ \ 
    \begin{pmatrix} b^{\rm out}_{{\rm L}\uparrow} \\
    e^{i \eta_{+} + i \phi/2} a^{\rm in}_{{\rm L}\uparrow} \\
    b^{\rm out}_{{\rm R}\uparrow} \\
    e^{i \eta_{+} - i \phi/2} a^{\rm in}_{{\rm R}\uparrow}
    \end{pmatrix} =\, S_{\rm F}^*
    \begin{pmatrix} b^{\rm in}_{{\rm L}\uparrow} \\
    e^{-i \eta_{+} + i \phi/2} a^{\rm out}_{{\rm L}\uparrow} \\
    b^{\rm in}_{{\rm R}\uparrow} \\
    e^{-i \eta_{+} - i \phi/2} a^{\rm out}_{{\rm R}\uparrow}
    \end{pmatrix}.
\end{equation}
The energy dependence of $S_{\rm F}$ has been dropped, since we work in the short-junction limit, see Eqs.~\eqref{eq:rHsimpl}. To first order in $\theta$, the normal-state scattering matrix $S_{\rm F}$ reads
\begin{equation}
  S_{\rm F} = \begin{pmatrix} 0 & 0 & e^{i k_0 L} & 0 \\ 0 & -1 & 0 & 0 \\ e^{i k_0 L} & 0 & 0 & 0 \\ 0 & 0 & 0 & -1 \end{pmatrix} +
  \frac{\theta}{2} \begin{pmatrix} 0 & 1 & 0 & e^{i k_0 L} \\ 1 & 0 & e^{i k_0 L} & 0 \\
  0 & e^{i k_0 L} & 0 & 1 \\ e^{i k_0 L} & 0 & 1 & 0 \end{pmatrix}.
  \label{eq:Stheta}
\end{equation}

For $\theta=0$, Eq.~\eqref{eq:aS} separates into disjoint equations for the electron-like and hole-like majority amplitudes.  For the
electron-like amplitudes one finds
\begin{equation}
    \begin{pmatrix}
        a^{\rm out}_{{\rm L}\uparrow}\\
        a^{\rm out}_{{\rm R}\uparrow}
    \end{pmatrix}
    =
    \begin{pmatrix}
        0 & e^{ik_0L}\\
        e^{ik_0L} & 0
    \end{pmatrix}
    \begin{pmatrix}
        a^{\rm in}_{{\rm L}\uparrow}\\
        a^{\rm in}_{{\rm R}\uparrow}
    \end{pmatrix},
    \qquad
    \begin{pmatrix}
        a^{\rm in}_{{\rm L}\uparrow}\\
        a^{\rm in}_{{\rm R}\uparrow}
    \end{pmatrix}
    =
    - e^{-2i\eta_+}
    \begin{pmatrix}
        a^{\rm out}_{{\rm L}\uparrow}\\
        a^{\rm out}_{{\rm R}\uparrow}
    \end{pmatrix}.
\label{eq:a_theta0}
\end{equation}
Similarly, for the hole-like amplitudes,
\begin{equation}
    \begin{pmatrix}
        b^{\rm out}_{{\rm L}\uparrow}\\
        b^{\rm out}_{{\rm R}\uparrow}
    \end{pmatrix}
    =
    \begin{pmatrix}
        0 & e^{-ik_0L}\\
        e^{-ik_0L} & 0
    \end{pmatrix}
    \begin{pmatrix}
        b^{\rm in}_{{\rm L}\uparrow}\\
        b^{\rm in}_{{\rm R}\uparrow}
    \end{pmatrix},
    \qquad
    \begin{pmatrix}
        b^{\rm in}_{{\rm L}\uparrow}\\
        b^{\rm in}_{{\rm R}\uparrow}
    \end{pmatrix}
    =
    - e^{-2i\eta_-}
    \begin{pmatrix}
        b^{\rm out}_{{\rm L}\uparrow}\\
        b^{\rm out}_{{\rm R}\uparrow}
    \end{pmatrix}.
\label{eq:b_theta0}
\end{equation}
From these conditions, it immediately follows that the coupled equations for the electron-like majority coefficients have a solution only if 
\begin{equation}
    e^{2 i k_0 L} = e^{4 i \eta_{+}(\tilde E)};
  \label{eq:Econd}
\end{equation}
while the equations for the hole-like majority coefficients have a nontrivial solution if
\begin{equation}
    e^{2 i k_0 L} = e^{-4 i \eta_{-}(\tilde E)}.
  \label{eq:Econdh}
\end{equation}
Since we consider an Andreev bound state that is electron-like for $\theta = 0$, we choose the energy $\tilde E$ such that the condition~\eqref{eq:Econd} is satisfied, while setting the hole-like majority coefficients to zero. In that case, one has
\begin{equation}
  \label{eq:esol}
    a^{\rm in}_{{\rm R}\uparrow} = \frac{1}{{\sqrt{{\cal N}_+}}},\ \
    a^{\rm out}_{{\rm R}\uparrow} = -\frac{e^{2 i \eta_+}}{{\sqrt{{\cal N}_+}}},\ \
    a^{\rm in}_{{\rm L}\uparrow} = - \frac{e^{2 i \eta_+ - i k_0 L}}{{\sqrt{{\cal N}_+}}},\ \
    a^{\rm out}_{{\rm L}\uparrow} = \frac{e^{4 i \eta_+ - i k_0 L}}{{\sqrt{{\cal N}_+}}} = \frac{e^{i k_0 L}}{{\sqrt{{\cal N}_+}}}
\end{equation}
for the electron-like coefficients, where ${\cal N}_+$ is a normalization factor, see Eq.~\eqref{eq:N_sigma}. The condition~\eqref{eq:Econd} gives the two solutions labeled by $j=1,2$. 

If $\theta$ is nonzero, the Andreev bound state also acquires nonzero hole-like majority amplitudes. To find the corresponding coefficients, we consider the coupled linear equations for the hole-like coefficients. These are inhomogeneous equations, in which the solution~\eqref{eq:esol} for the electron-like coefficients, multiplied by the second term in Eq.~\eqref{eq:Stheta}, appears as a source term. Eq.~\eqref{eq:bRjout} follows upon solving this inhomogeneous linear set. The denominator $\sin(2(\zeta+\eta_-))$ in Eq.~\eqref{eq:B_j} reflects that a bounded solution for the hole-like coefficients exists only if this set of equations is non-singular. This is the case unless Eq.~\eqref{eq:Econdh} is satisfied, which is precisely when there is a hole-like Andreev bound state at energy $E$.

\section{S/HM/S Junction: current carried by the continuum spectrum \label{app:SHMS}}

In this appendix, we provide the explicit expression for the contribution $j_{\sigma}(\tilde E)$ from the continuous part of the spectrum for small tilt angle $\theta$. We first consider energies $|\Delta| - (1/2) \hbar |\Omega| < \tilde E < |\Delta| + (1/2) \hbar |\Omega|$. In this energy range, there is a contribution from scattering states with $\sigma = \mbox{sign}\, \Omega$ only. Away from resonances, one has
\begin{equation}
  j_{\sigma}(\tilde E) = \theta^2 j'_{\sigma}(\tilde E),
\end{equation}
up corrections sub-leading in $\theta$, with
\begin{align}
  j'_\sigma(\tilde E) =&\, \delta_{\sigma,{\rm sign}\,\Omega}
  \frac{i \sin (\eta_{\sigma}) \left[\sin(\sigma k_0 L) \cos(2 \eta_{\sigma}) \sin \eta_{-\sigma} -\cos (k_0 L) \cos \eta_{\sigma} (\sin^2 \eta_{\sigma} + \sin^2 \eta_{-\sigma}) \right]}{\pi \sin(\sigma k_0 L + 2 \eta_{-\sigma}) (\cos^2(k_0 L) - \cos^2(2\eta_\sigma))}
\end{align}
Near a resonance, for small $\theta$, $j_{\sigma}(\tilde E)$ is given by Eq.\ (\ref{eq:Jresonance}) with
\begin{align} 
  J_{\alpha} =&\, \lim_{\tilde E \to \tilde E_{\alpha}} (\tilde E - \tilde E_{\alpha})
  j_{\sigma}(\tilde E), \label{eq:app:J} \\
    \Gamma_\alpha =&\, - \lim_{\tilde E \to \tilde E_\alpha} (\tilde E - \tilde E_\alpha) \frac{\cos(2\bar\eta)\cos(2\eta') + \cos(2Lk_0 -2\bar \eta) - \cos(2[Lk_0 - 2\bar\eta]) -\cos(2\eta') - 4 \cos(\phi) \sin^2\bar\eta \cos^2 \eta'}{2 \sin(\sigma Lk_0 + 2\eta_{-\sigma})\sin(\sigma Lk_0 - 2\eta_{\sigma})}, \label{eq:app:Gamma}
\end{align}
and
\begin{gather}
  \delta_\alpha = \lim_{\tilde E \to \tilde E_\alpha} (\tilde E - \tilde E_\alpha)\frac{n(\tilde E)}{d(\tilde E)}.
  \label{eq:delta}
\end{gather}
In Eq.\ (\ref{eq:delta}) the numerator and denominator are given by the following lengthy expressions:
\begin{align}
    n(\tilde E) = & 3 \sin\!\left(2 \sigma k_0 L - \eta_{-\sigma}\right) -
    (9 + 2 \cos(2 \eta_{\sigma})) \sin \eta_{-\sigma}
    - (3 + 4 \cos(2 \eta_{\sigma})) \sin(3 \eta_{-\sigma})
    - 2 \cos(3 \eta_{\sigma})
    \bigl(
    \sin(2 \eta_{-\sigma}) \nonumber \\
    & + 4 \sin\!\left(2\sigma k_0 L + 2 \eta_{-\sigma}\right)
    \bigr) - 2 (6 + 5 \cos(2 \eta_{\sigma})) \sin\!\left(2 \sigma k_0 L + \eta_{-\sigma}\right)
    + 2 \cos \eta_{\sigma} \Bigl(
    2 \sin(2 \sigma k_0 L)
    + 9 \sin(2 \eta_{-\sigma}) \nonumber \\
    & - \sin(4 \eta_{-\sigma})
    + 6 \sin\!\left(2(\sigma k_0 L + \eta_{-\sigma})\right)
    - 4 \sin\!\left(2\sigma k_0 L + 4 \eta_{-\sigma} \right)
    \Bigr)
    + (3 + 16 \cos(2 \eta_{\sigma})) \sin\!\left(2 \sigma k_0 L + 3 \eta_{-\sigma}\right), \\
    d(\tilde E) = & 48 \sin(\sigma L k_0 + 2\eta_{-\sigma}) \big[\sin(\sigma L k_0) \cos(2\eta_{\sigma})\sin \eta_{-\sigma} -\cos(L k_0)\cos \eta_\sigma (\sin^2 \eta_{\sigma} + \sin^2\eta_{-\sigma})\big].
\end{align}
The apparent singularity associated with the factor $\sin(\sigma Lk_0 + 2\eta_{-\sigma})$ in the denominator is regularized by the prefactor $(\tilde E - \tilde E_\alpha)$, so that $\Gamma_\alpha$ and $\delta_\alpha$ remain finite when taking the limit $\tilde E \to \tilde E_\alpha$.
Since in the final expression for the current, only the integral of $j_{\sigma}(\tilde E)$ over $\tilde E$ appears, for $|\tilde E - \tilde E_{\alpha}| \ll |\Delta|$, $|\Omega|$ we may approximate
\begin{equation}
  j_{\sigma}(\tilde E) = {\rm pv}\, \theta^2 j'_{\sigma} (\tilde E) +
  \pi \theta^2 \sum_{\alpha}  \delta(\tilde E - \tilde E_{\alpha})
  \frac{(\delta_{\alpha} - \mbox{Re}\, \Gamma_{\alpha})}
       {|\mbox{Im}\, \Gamma_{\alpha}|},
  \label{eq:jdelta}
\end{equation}
where ``pv'' denotes the principal value and any correction terms are sub-leading in the limit $\theta \ll 1$. Substituting the explicit expressions given above, we may then write $j_{\sigma}(\tilde E)$ as
\begin{align}
  \label{eq:jE}
  j_{\sigma}(\tilde E) = & \delta_{\sigma,{\rm sign}\, \Omega} 
  \theta^2 \sin \phi
  \left[ 
  {\rm pv}\, C_0^\sigma + C_1^\sigma \delta(\sin(\sigma k_0 L + 2 \eta_{-\sigma})) \right],
\end{align} 
where 
\begin{align}
  \label{eq:CE}
  C_0^\sigma =&\,
  i \sin \eta_{\sigma} 
  \frac{\sin(\sigma k_0 L) \cos(2 \eta_{\sigma}) \sin \eta_{-\sigma}
  -\cos (k_0 L) \cos \eta_{\sigma} 
  (\sin^2 \eta_{\sigma} + \sin^2 \eta_{-\sigma}))}
  {\pi \sin(\sigma k_0 L + 2 \eta_{-\sigma}) \sin^2(\sigma k_0 L - 2 \eta_{\sigma})}, \nonumber \\  
  C_1^\sigma =&\,
  \cos(k_0L+2\eta_{-\sigma})\sin \eta_{-\sigma}
  \frac{1- \cos(2\eta_{\sigma})\sin^2\eta_{-\sigma}
    +\cos\eta_{\sigma}(\cos(2\eta_{\sigma}) - 2)\cos\eta_{-\sigma}}
  {\cos(3\eta_{\sigma})-2\cos\eta_{-\sigma}(\cos(2\eta_{\sigma})+\cos(2\eta_{-\sigma}))
  + \cos\eta_{\sigma}(1+2\cos(2\eta_{-\sigma}))}.
\end{align}

For $\tilde{E} > |\Delta| + \hbar |\Omega|/2$, both spin projections of quasiparticles can directly enter the continuum spectrum in the adjacent superconductors and one finds
\begin{align}
    j_\sigma(\tilde{E}) =&\, - i \frac{\sigma \theta^2}{\pi}
\frac{\sin(2\eta_{+})\sin(2k_0 L)\sin (2\bar \eta) \sin (2\eta') \sin \phi }{(\cos(2k_0 L) - \cos(4\eta_+))(\cos(2k_0 L) - \cos(4\eta_-))}.
\end{align}
When $k_B T \ll \hbar |\Delta|$, the difference between $f_{\uparrow}(\tilde E)$ and $f_{\downarrow}(\tilde E)$ becomes small for $\tilde E > |\Delta|$, so that the two contributions from $\sigma = \uparrow$, $\downarrow$ cancel each other and do not contribute to the current.

\section{S/HM/S Junction: non-trivial $\hbar\Omega \ll |\Delta|$ asymptotics \label{app:SHMS:Omega<<Delta}}

In this appendix, we give an example of a 1D S/HM/S Josephson junction for which the $\Omega^2$ dependence breaks down when $\hbar\Omega \ll |\Delta|$ and at $T = 0$. We consider a junction satisfying $\zeta \ll \hbar\Omega/|\Delta| \ll 1$, where $\Omega > 0$ and $\zeta$ is defined in Eq.~\eqref{eq:HM:zeta}. The bound-state spectrum is then close to that shown in the top panel of Fig.~\ref{fig:ABS_HM}, near the points $k_0L = 0, \pi, \dots$. In this regime, $\tilde E_{{\rm e}1}$ and $\tilde E_{{\rm h}2}$ contribute to the current, and their respective contributions, see Eq.~\eqref{eq:dEdphi}, are given by
\begin{align}
    \frac{\partial \delta \tilde E_{{\rm e}1}}{\partial \phi} & = \theta^2 \sin \phi |\Delta| \zeta \frac{\sqrt{\hbar \Omega/|\Delta|}}{4\sqrt{2}},  \\
    \frac{\partial \delta \tilde E_{{\rm h}2}}{\partial \phi} & = \theta^2 \sin \phi |\Delta| \frac{(\hbar \Omega)^3}{32 |\Delta|^3}.
\end{align}
On the other hand, the current carried by the continuum spectrum with energies $|\Delta| - (1/2) \hbar |\Omega| < \tilde E < |\Delta| + (1/2) \hbar |\Omega|$ can be found to be
\begin{gather}
    I_{\rm cont} = \theta^2 \sin\phi |\Delta| \zeta \frac{\sqrt{\hbar \Omega/|\Delta|}}{4\sqrt{2}} \frac{\ln(2^{-3} e^{8+\pi} [\zeta/\sqrt{(\hbar \Omega/|\Delta|)}]^2)}{2\pi}.
\end{gather}
Taken together, this provides a clear example in which $I_{\rm c}$ exhibits a nontrivial, i.e. non-quadratic, dependence on $\Omega$.

\section{S/F/S Junction: current carried by the continuum spectrum \label{app:SFS}}
For the continuum part of the spectrum, we find that the current carried by a state at energy $\tilde{E}$, to leading order in $\theta$, is given by
\begin{gather}
    j_\sigma(\tilde E) = \sigma \sum_{s = \pm} 2is \frac{\sin(2\eta_{\sigma}) \sin^2(\zeta_{-s\sigma} + \sigma \eta_{-\sigma}) + \theta^2 \sin\eta_{\sigma} [\sin^2 \kappa L \cos(\eta_{-\sigma} - s\phi) - \sin \kappa L \cos\eta_\sigma \sin(\kappa L - s\sigma\phi + 2 \sigma\eta_{-\sigma})]}{\left|2 \sin(\zeta_{-s\sigma} + \sigma \eta_{-\sigma}) \sin(\zeta_{s\sigma} + \sigma \eta_{-\sigma}) + \theta^2 [\sin^2\kappa L - \sin (\sigma\kappa L) \sin(\sigma \kappa L + \eta_{\sigma} + \eta_{-\sigma})]\right|^2}.
\end{gather}
Setting $\theta = 0$, we retrieve Eq.~\eqref{eq:SFS:continuum} of the main text. Retaining corrections up to quadratic order in $\theta$ in the numerator and denominator yields the leading correction to the supercurrent $\delta j_\sigma(\tilde E)$, which contains the contribution from equal-spin Cooper pairs.

\section{Details of numerical calculation of the current \label{app:numerics}}

\begin{figure*}[t]
\centerline{ \includegraphics[width=.9\textwidth]{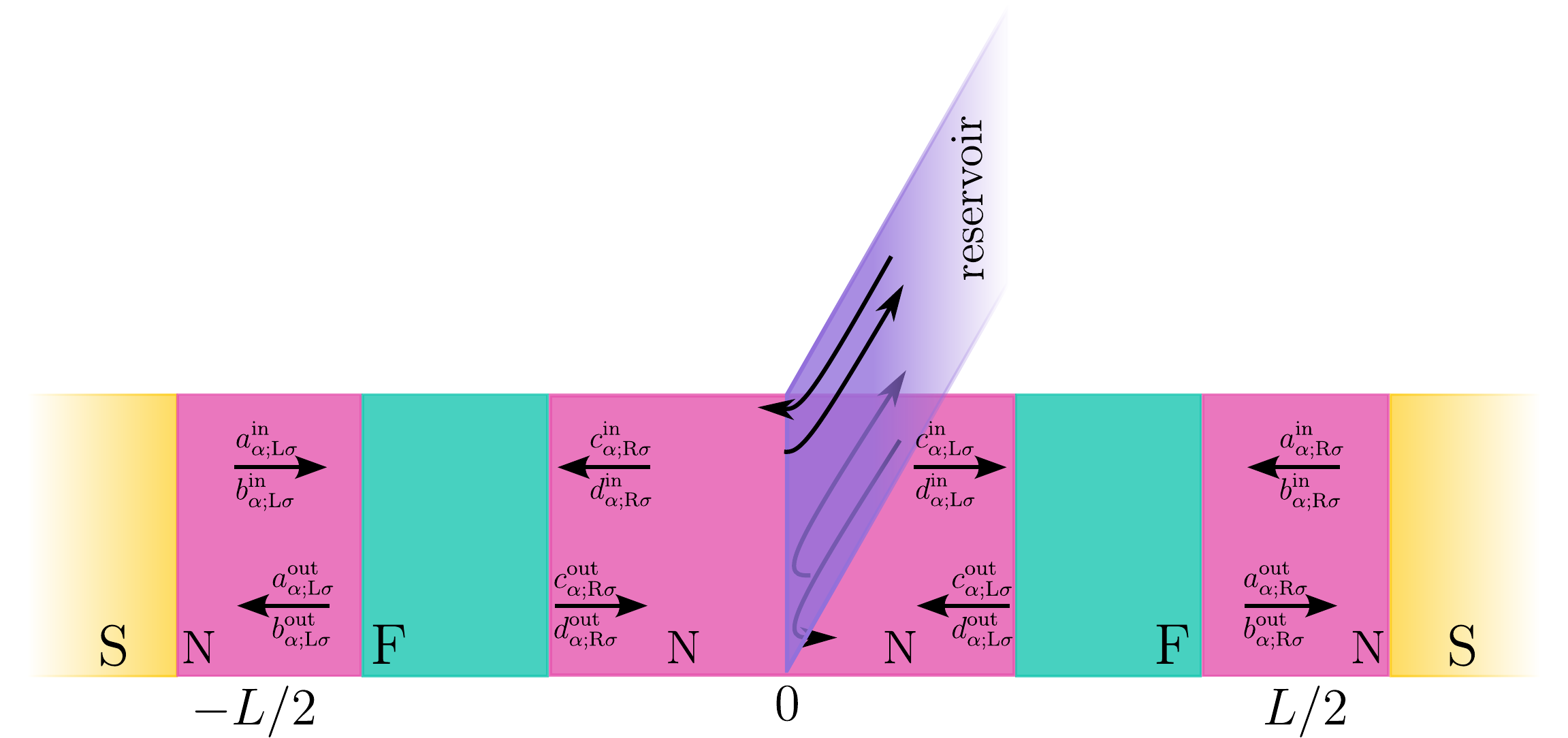}}
\caption{Geometry used for the numerical calculation of the current in the junction. A normal reservoir in thermal equilibrium is weakly coupled to an ideal lead inserted at $x=0$, at the center of the magnetic region. The amplitude that electrons scatter from the ideal lead into the normal reservoir or vice versa is $i \sqrt{\Gamma}$, see Eq.\ (\ref{eq:R}). The coupling to the normal lead imposes the population (\ref{eq:nalpha}) for the Andreev bound states in the junction and turns them into scattering resonances. In our numerical calculations, we set $\Gamma = (0.05)^2$. We verify that the small level broadening due to the coupling to the normal reservoir does not affect our results.
\label{fig:Appendix}} 
\end{figure*}

In this appendix, we discuss the implementation of the numerical evaluation of the Josephson current. While Eq.~\eqref{eq:IEdiscrete} is instructive, because it highlights the contribution of the Andreev bound states to the supercurrent and, hence, explains the sharp jumps of the supercurrent with system parameters in terms of discontinuous related to the population of Andreev levels, the separation into discrete- and continuous-spectrum contributions to the supercurrent poses a complication from a numerical point of view. These complications can be avoided by weakly coupling the central magnetic part of the junction to a normal reservoir, as illustrated in Fig.~\ref{fig:Appendix}. 

In the presence of the additional normal reservoir, solutions of the Bogoliubov-de Gennes equation (\ref{eq:HBdGEq}) are scattering states with a continuous spectrum for all energies $\tilde E$, so that a separate treastment of discrete and continuous parts of the spectrum is no longer necessary. With the additional normal lead, the Andreeev bound states turn into scattering resonances. The additional lead acts as a source of thermalization for the Andreev bound states, populating them according to Eq.~\eqref{eq:nalpha}. The resonance width is proportional to the transmission probability $\Gamma$ between the modes in the reservoir and in the junction. In our numerical calculations, we use $\Gamma = (0.05)^2$, which enables us to resolve these resonances while still ensuring that the coupling to the additional lead has infinitesimal effect on the Josephson current itself. 

To implement the coupling to the additional normal reservoir, we insert an additional ideal lead at the center of the magnetic region, {\em i.e.}, at coordinate $x = 0$. Like the other two ideal leads, the length of this ideal lead is taken to zero at the end of the calculation. Left- and right-moving states in the ideal lead around $x=0$ are coupled separately to modes in an additional lead, which is connected to the normal reservoir, see Fig.\ \ref{fig:Appendix}.

The addition of the normal reservoir requires that the composite index $\alpha = (\tilde E,\beta,l,\tau,\sigma)$ labeling the scattering states now has five elements, whereby the additional element $l = 0$ for scattering states incident from one of the superconductors and $l=1$ for scattering states incident from the normal reservoir. For states incident from the normal reservoir, we set $\beta = 1$ (${\rm R}$) if it couples to a left-moving state in the ideal lead (which has amplitudes $c^{\rm in}_{\alpha;{\rm R}\sigma}$ and $c^{\rm out}_{\alpha;{\rm L}\sigma}$ for electron-like states and $d^{\rm in}_{\alpha;{\rm R}\sigma}$ and $d^{\rm out}_{\alpha;{\rm L}\sigma}$ for hole-like states, see Fig.\ \ref{fig:Appendix}) and $\beta = -1$ (${\rm L}$) if it couples to a right-moving state in the ideal lead (corresponding to the amplitudes $c^{\rm in}_{\alpha;{\rm L}\sigma}$ and $c^{\rm out}_{\alpha;{\rm R}\sigma}$, $d^{\rm in}_{\alpha;{\rm L}\sigma}$, and $d^{\rm out}_{\alpha;{\rm R}\sigma}$).

Referring to Fig.~\ref{fig:Appendix} for the geometry and the notation of the amplitudes of the scattering states in the additional ideal lead around $x=0$, we replace Eqs.\ (\ref{eq:SF}) and (\ref{eq:SFh}) by the following system of equations:
\begin{align}
  \label{eq:F1}
  \begin{pmatrix}
    a_{\alpha;{\rm L}\uparrow}^{{\rm out}} \\
    a_{\alpha;{\rm L}\downarrow}^{{\rm out}}\\
    c_{\alpha;{\rm R}\uparrow}^{{\rm out}} \\
    c_{\alpha;{\rm R}\downarrow}^{{\rm out}}
  \end{pmatrix}
  =&\, S_{\rm F}(\tilde E)
  \begin{pmatrix}
    a_{\alpha;{\rm L}\uparrow}^{{\rm in}} \\
    a_{\alpha;{\rm L}\downarrow}^{{\rm in}}\\
    c_{\alpha;{\rm R}\uparrow}^{{\rm in}} \\
    c_{\alpha;{\rm R}\downarrow}^{{\rm in}}
  \end{pmatrix},& \quad \begin{pmatrix}
    b_{\alpha;{\rm L}\uparrow}^{{\rm out}} \\
    b_{\alpha;{\rm L}\downarrow}^{{\rm out}}\\
    d_{\alpha;{\rm R}\uparrow}^{{\rm out}} \\
    d_{\alpha;{\rm R}\downarrow}^{{\rm out}}
  \end{pmatrix}
  =&\, S_{\rm F}(-\tilde E)^*
  \begin{pmatrix}
    b_{\alpha;{\rm L}\uparrow}^{{\rm in}} \\
    b_{\alpha;{\rm L}\downarrow}^{{\rm in}}\\
    d_{\alpha;{\rm R}\uparrow}^{{\rm in}} \\
    d_{\alpha;{\rm R}\downarrow}^{{\rm in}}
  \end{pmatrix}, \\
  \label{eq:F2}
  \begin{pmatrix}
    c_{\alpha;{\rm L}\uparrow}^{{\rm out}} \\
    c_{\alpha;{\rm L}\downarrow}^{{\rm out}}\\
    a_{\alpha;{\rm R}\uparrow}^{{\rm out}} \\
    a_{\alpha;{\rm R}\downarrow}^{{\rm out}}
  \end{pmatrix}
  =&\, S_{\rm F}(\tilde E)
  \begin{pmatrix}
    c_{\alpha;{\rm L}\uparrow}^{{\rm in}} \\
    c_{\alpha;{\rm L}\downarrow}^{{\rm in}}\\
    a_{\alpha;{\rm R}\uparrow}^{{\rm in}} \\
    a_{\alpha;{\rm R}\downarrow}^{{\rm in}}
  \end{pmatrix},& \quad \begin{pmatrix}
    d_{\alpha;{\rm L}\uparrow}^{{\rm out}} \\
    d_{\alpha;{\rm L}\downarrow}^{{\rm out}}\\
    b_{\alpha;{\rm R}\uparrow}^{{\rm out}} \\
    b_{\alpha;{\rm R}\downarrow}^{{\rm out}}
  \end{pmatrix}
  =&\, S_{\rm F}(-\tilde E)^*
  \begin{pmatrix}
    d_{\alpha;{\rm L}\uparrow}^{{\rm in}} \\
    d_{\alpha;{\rm L}\downarrow}^{{\rm in}}\\
    b_{\alpha;{\rm R}\uparrow}^{{\rm in}} \\
    b_{\alpha;{\rm R}\downarrow}^{{\rm in}}
  \end{pmatrix},
\end{align}
Here Eqs.\ (\ref{eq:F1}) and (\ref{eq:F2}) describe scattering of electron ($a$ and $c$) and hole ($b$ and $d$) components, see Fig.~\ref{fig:Appendix}, around the left and right parts of the central F region ($-L/2 < x < 0$ and $0 < x < L/2$), respectively.
The relations between scattering amplitudes at the superconductor interfaces are still given by Eqs.~\eqref{eq:SS} and~\eqref{eq:SScontinuous} of the main text, whereby the source terms in Eq.\ (\ref{eq:SScontinuous}) acquire an additional factor $\delta_{l0}$. The set of equation is closed by the relations between the amplitudes on the left and right sides of the normal reservoir (see Fig.\ \ref{fig:Appendix}),
\begin{align} 
  c_{\alpha;\beta',\sigma'}^{{\rm in}} =&\,
  \sqrt{1 - \Gamma}
  c_{\alpha;-\beta',\sigma'}^{{\rm out}}
  + i \sqrt{\Gamma} \delta_{\beta\beta'} \delta_{\sigma\sigma'} \delta_{l1} \delta_{\tau{\rm e}}, \nonumber \\
  d_{\alpha;\beta',\sigma'}^{{\rm in}} =&\,
  \sqrt{1 - \Gamma}
  d_{\alpha;-\beta',\sigma'}^{{\rm out}}
  - i \sqrt{\Gamma} \delta_{\beta\beta'} \delta_{\sigma\sigma'} \delta_{l1} \delta_{\tau{\rm h}},
  \label{eq:R}
\end{align}
The current is calculated using Eqs.\ (\ref{eq:ItotalRL}) and (\ref{eq:Itotal}) from the main text, where there summation over the composite index $\alpha$ should be understood as $\sum_{\beta,l,\sigma,\tau} \int d\tilde E \ldots$.

The source code for the numerical calculations can be found in the Zenodo repository~\cite{andriyakhina2026}.

\end{widetext}

\end{document}